\documentclass[10pt, aps, prc, superscriptaddress, nofootinbib, 
               amsmath, 
               twocolumn, preprintnumbers, 
               showpacs, 
               raggedbottom, 
               floatfix, 
               longbibliography, 
               colorlinks=true, linkcolor=blue, citecolor=blue, urlcolor=blue, 
              ]{revtex4-1}
\pdfoutput=1
\usepackage[T1]{fontenc}
\usepackage[utf8]{inputenc}
\usepackage[secnoindent, 
            ]{my_paper} 
\usepackage{my_acronyms}        
\usepackage[english]{babel}     
\usepackage{adjustbox}          
\usepackage{colortbl}
\usepackage{xparse}             
\usepackage[caption=false]{subfig}
\usepackage[percent]{overpic} 

\usepackage[
            tightpage]{preview}

\renewcommand{\vect}[1]{\boldsymbol{#1}}

\newcommand{\n}{n}  

\newcommand{\beq}{\begin{equation}}
\newcommand{\eeq}{\end{equation}}
\newcommand{\bea}{\begin{eqnarray}}
\newcommand{\eea}{\end{eqnarray}}

\begin{document}

\newlength{\mybaselineskip}
\setlength{\mybaselineskip}{\baselineskip}

\title{Fission Dynamics  of $^{240}$Pu from Saddle-to-Scission and Beyond}

\author{Aurel Bulgac}%
\affiliation{Department of Physics,  University of Washington, Seattle, Washington 98195--1560, USA}

\author{Shi Jin}%
\affiliation{Department of Physics,  University of Washington, Seattle,  Washington 98195--1560, USA}
  
\author{Kenneth J. Roche}%
\affiliation{Pacific Northwest National Laboratory, Richland, WA 99352, USA}
\affiliation{Department of Physics,  University of Washington, Seattle,  Washington 98195--1560, USA}

\author{Nicolas Schunck}%
\affiliation{Nuclear and Chemical Science Division, 
  Lawrence Livermore National Laboratory, Livermore,   CA 94551, USA}

\author{Ionel Stetcu}%
\affiliation{Theoretical Division, Los Alamos National Laboratory, Los Alamos, NM 87545, USA}

\date{\today}

\begin{abstract}

Calculations are presented for the time evolution of $^{240}$Pu from
the proximity of the outer saddle point until the fission fragments
are well separated, using the time-dependent density functional theory
extended to superfluid systems.  We have tested three families of
nuclear energy density functionals and found that all functionals exhibit
a similar dynamics: the collective motion is highly dissipative and with
little trace of inertial dynamics, due to the one-body dissipation
mechanism alone. This finding justifies the validity of using the
overdamped collective motion approach and to some extent the main
assumptions in statistical models of fission.  This conclusion is
robust with respect to the nuclear energy density functional used. The
configurations and interactions left out of the present theory
framework only increase the role of the dissipative couplings. An
unexpected finding is varying the pairing strength within a quite large
range has only minor effects on the dynamics.  We find notable
differences in the excitation energy sharing between the fission
fragments in the cases of spontaneous and induced fission. 
With increasing initial excitation energy of the fissioning nucleus
more excitation energy is deposited in the heavy fragment, in
agreement with experimental data on average neutron multiplicities.   

\end{abstract}

\preprint{NT@UW-18-03, LA-UR-18-23917}

\glsreset{NEDF}
\glsreset{GDR}
\glsreset{DFT}
\glsreset{UFG}

\maketitle


\glsreset{NEDF}
\glsreset{GDR}
\glsreset{DFT}
\glsreset{UFG}

\section{Introduction} \label{sec:I}

Eighty years after the
discovery of nuclear fission~\cite{Hahn:1939} a full microscopic
description is still lacking, which in itself is perhaps a world
record in quantum many-body theory.  The term nuclear fission was 
coined by Meitner~\cite{Meitner:1939,Jorgensen:2019}. In 1934 Ida
Noddack~\cite{Noddack:1934} presented credible arguments that perhaps
Enrico Fermi~\cite{Fermi1:1934} had already created fission fragments
in his laboratory. Fermi had bombarded uranium with neutrons, but
failed to observe the fission fragments by shielding his uranium
target with a thin aluminum foil in order to minimize the background due to
$\alpha$-particles~\cite{Fermi2:1934}, which likely blocked the
fission fragments too~\cite{Pearson:2015}.  Reasoning based on the Gamow
theory of quantum tunneling led many at the time to expect that
fission would occur on time scales many orders of magnitude longer
than the age of the Universe.  This explains the shock experienced by
the scientific community when Hahn and Strassmann published their
observations on January 6th, 1939 (submitted on December 22nd,
1938 and unfortunately without Meitner as a coauthor)~\cite{Hahn:1939}.  
\textcite{Meitner:1939}, who became aware of
these results during the last days of 1938, figured out the basic
explanation of nuclear fission even before the Hahn and Strassmann
paper appeared in print.  They presented compelling arguments that
Gamow's 1930 charged liquid drop model of
nuclei~\cite{Gamow:1930,Stuewer:2010}, in which the Coulomb
interaction between protons competes with the surface nuclear tension,
leads to a very natural explanation of the main fission
properties. The liquid drop model was almost immediately combined with
Bohr's compound model and extended to deformed nuclei by Bohr and
Wheeler~\cite{Bohr:1939}. According to Bohr and
Wheeler~\cite{Bohr:1939}, a low energy incident neutron is captured by
the uranium nucleus and leads to the formation of a compound
nucleus~\cite{Bohr:1936}. For example, the energy levels in a compound nucleus are separated by
$\Delta E\approx 10$ eV in the $^{232}$Th+n
reaction, and with similar order of magnitude 
in  heavy nuclei~\cite{Bohr:1969,RIPL:3,Egidy:2005}. Thus the evolution of the nuclear shape
from a rather compact one, corresponding to the ground state of uranium after the neutron capture,
until it reaches the fission barrier lasts a relatively long time
the order ${\cal O}\left (\tfrac{\hbar}{\Delta E}\right)=0.6 \times10^{-16}\;\text{sec.}=2\times
10^7$ fm/c. This time is much longer than the time needed for a
nucleon to traverse a nucleus back and forth, which is approximately
$1.7\times 10^{-22}\; \text{sec.} = 50$ fm/c. As a result the memory of the initial 
state it is ``forgotten'' and statistical arguments can be use to describe 
the eventual decay of a compound nucleus and its decay 
various branching ratios~\cite{Weisskopf:1937,Hauser:1952}.
The position of the
fission barrier is determined by the nuclear elongation, where the rate of 
increase of the nuclear surface energy is exactly compensated by the rate of 
decrease of the Coulomb energy of the nucleus.  Since the role of the
shell-effects and the formation of the fission isomer second well were
understood only much later~\cite{Strutinsky:1967,BRACK:1972,Bjornholm:1980}, Bohr and
Wheeler could not tackle the asymmetric fission and theoretically addressed
only the case of symmetric fission. After reaching the outer
fission barrier a nucleus evolves towards the scission configurations
into two separated fission fragments (FFs) at a much faster rate. During this
non-equilibrium evolution of the mother nucleus from saddle-to-scission 
the properties of the FFs are defined. 

Since an accurate solution of the time-dependent Schr\"odinger
equation with realistic nucleon interactions will be out of reach for
a very long time (if ever), the question arises: what would be a
reliable microscopic approach?  A  Feyman's
real-time path integral formulation~\cite{Negele:1988,Bulgac:2010} of
quantum many-body systems is particularly appealing.  The many-body
wave function is represented as a sum over all possible paths joining the
initial and final configurations, with appropriate weights:  
\begin{equation}
\!\!\!\!\!\Psi(t) =
\int \mathcal{D}[\sigma(t)] W[\sigma(t)] 
\exp\left( -\frac{i}{\hbar}\int_{t_{i}}^{t_{f}} \hat{h}[\sigma(t)]\right)
\Psi(0),  \label{eq:pathint}
\end{equation}
here $\mathcal{D}[\sigma(t)]$ is an appropriate measure depending on all 
auxiliary fields, $W[\sigma(t)]$ is a Gaussian weight and 
$\hat{h}[\sigma(t)]$ is a one-body Hamiltonian built with the auxiliary one-body 
fields $\sigma(t)$. $\Psi(0)$ is the initial wave function, often chosen as a 
(generalized) Slater determinant.
\footnote{In reality the initial wave function is an ensemble of many Slater determinants, and while each 
member of the ensemble might break various symmetries, the total wave function 
satisfies all symmetries. Typically a fissioning nucleus in its intrinsic ground state 
has a quadrupole deformation and positive parity for example. However on the way to the saddle 
this state evolves into one with a non-vanishing octupole moment, which at first glance 
appears to be an impossible transition. In large many-body systems however one observes 
remnants of the spontaneous symmetry breaking, which strictly speaking exists only in 
infinite systems. The density of these states is however so large that the time for a 
nucleus to ``tunnel'' from one symmetry breaking state to another, in order to restore 
the symmetry, is much larger that the time it takes a nucleus to evolve, for example 
from a state with positive octupole momentum $Q_{30}>0$ (in center-of-mass reference 
frame)  to one with negative octupole
momentum  $Q_{30}<0$, which in the long run will restore the parity. The large time-difference 
in the scales of the processes which are responsible for the restoration of symmetries 
and the time scale of the fission dynamics in our case, allows us to 
focus on the dynamics of a single component of the ensemble at a time. We thank J. Randrup 
for urging us to shed light on this issue.}

Thus, the 
true many-nucleon wave function is now a time-dependent linear superposition of 
many time-dependent (generalized) Slater determinants. In this respect the true
many-nucleon wave function has a similar mathematical structure as the wave 
function in the time-dependent generator coordinate method (TDGCM) introduced 
by Wheeler {\it et al.}~\cite{Hill:1953,Griffin:1957}, see \cref{sec:II}.
One  cannot fail but see here also the analogy in treating fluctuations around the mean field 
trajectory with the classical Langevin description of nuclear collective motion 
as well~\cite{Frobrich:1998}. 
The representation  \eqref{eq:pathint} (which is an exact one)  
of the many-body wave function has the 
great advantage that
each trajectory is independent of all the others.
 In the stationary phase
approximation such a path integral selects a particular
mean field, which can be interpreted as the most probable trajectory. This mean-field trajectory
is not uniquely defined~\cite{Negele:1988}, and fluctuations around it 
are important. The current attitude in nuclear physics, even tough usually not
explicitly spelled out,   
is to simulate this particular path with a trajectory generated in
time-dependent density functional theory (TDDFT). 

Even though the mean paths along which nuclei evolve do not convey the
whole story in fission, they do determine the most probable properties of
this non-equilibrium quantum process. A complete microscopic
characterization of the fission dynamics is still lacking, since
practically all simulations performed so far have relied on a range of
simplifying assumptions, the accuracy of which have not or could not have
been tested.  Here we will consider only the most probable fission
trajectories and leave the study of the role of fluctuations in a
fully quantum mechanical formulation to future studies~\cite{Bulgac:2019a}.

Many basic questions remained unanswered by the microscopic theory 
and experiment provides often only indirect and hard to
quantify insight.  What is the nature of the driving force in fission
dynamics?  What is the mechanism that provides excitation energy to FFs at scission?
How is this excitation energy between the FFs shared? 
Are the one-body~\cite{Blocki:1978} or/and the two-body excitation mechanisms effective?
Are pairing correlations still important in the later stages of
the evolution before scission? How many neutrons (if any) are emitted before or/and at
scission or/and before the fission products are fully accelerated? Are
ensembles of initial conditions important in modeling the FFs yields
and properties?  How the initial excitation energy of the fissioning nucleus 
impacts the excitation energy mechanism between FFs?
How the average neutron multiplicities as a function  of the FF mass 
are affected by  the 
initial excitation energy of the fissioning system?
The present study is our attempt to shed light on all
these questions.

\section{Main theoretical approaches to fission dynamics} \label{sec:II}

The evolution of the nuclear shape from the ground state to the outer
fission barrier is very slow and one often invokes the picture of an
adiabatic evolution, particularly in the case of spontaneous fission.
The description of the nuclear dynamics, starting when the nucleus
exits or passes the outer fission barrier until it reaches the
scission configuration, is treated in the literature either as an
adiabatic evolution (leading to a conservative dynamics) 
or as a damped or even over-damped motion, using the same kind of parameters.
There is no consensus in literature on the character of the dynamics
during this last phase of fission, namely conservative or Hamiltonian versus dissipative 
dynamics or even over-damped motion.
All different approaches, 
based on such different assumptions about the character of the fission dynamics
lead reasonably accurate  agreement with experiment. So
far, it has been impossible to observe directly in experiments this
stage of the nuclear dynamics.  Yet,  the most important properties of the FFs, 
their masses and charges, their shapes and intrinsic excitation energies, are defined
during this stage. 

One class of microscopic theoretical models used to describe
fission fragment yields is based on the time-dependent generator
coordinate method (TDGCM)~\cite{Hill:1953,Griffin:1957,Pomorski:2012,
Schunck:2016,Goutte:2005,Regnier:2016,Zdeb:2017,Schunck:2019,Younes:2019} or on the adiabatic 
time-dependent Hartree-Fock  (ATDHF) theory~\cite{Baranger:1972,Baranger:1978,Villars:1978,Ring:2004}.
It was establish almost forty years ago that TDGCM with complex coordinates 
is basically equivalent to ATDHF approach to large amplitude collective motion (LACM)~\cite{Goeke:1980} 
and for that reason we will concentrate here on TDGCM alone. 
The wave function of a many-fermion system in TDGCM is constructed according to the following prescription
\beq
\Psi({\bf x},t) =\int d{\bf q}  f({\bf q},t)\Phi({\bf x}|{\bf q}) ,\label{eq:GCM}
\eeq
where $\Phi({\bf x}|{\bf q})$ are (generalized) Slater determinants depending on nucleon spatial 
coordinates, spin, and isospin ${\bf x} =(x_1,\ldots,x_A),\; x_k=( {\bf r}_k, \sigma_k,\tau_k)$ and
parameterized by the collective coordinates ${\bf q}=(q_1,\ldots,q_n)$, and where $f({\bf q},t)$    
is the collective wave function. 

The collective coordinates can and are often interpreted as 
real degrees of freedom (DoF), which one assumes that can be decoupled from the rest, or the intrinsic DoF. 
The definition of collective and intrinsic DoF is still not a solved problem and it is not obvious 
even that a satisfactory solution even exists~\cite{Dang:2000}. An alternative interpretation it to treat treat 
these ``coordinates'' as mere labels of the (generalized) Slater determinants. In that case one can interpret 
${\bf q}$ as labels of ``sites'' from to and where to the nucleus hops 
during evolution, a model which appears to be much simpler and almost as accurate as the GCM~\cite{Bertsch:1991}, 
in manner analogous to the  tight-binding models in condensed matter theory~\cite{Mermin:1976}. 
In the case when one takes at face value that ${\bf q}$ are real collective coordinates the interpretation of the 
results in terms of ``real collective'' DoF could lead to inconsistencies of the 
emerging models and physical interpretation of the results. A particular source of difficulties 
lies in the fact that the total number of DoF in GCM $A+n$ is unphysical. 

The representation \eqref{eq:GCM} would be in principle exact if the (generalized) Slater determinants $\Phi({\bf x}|{\bf q})$ 
would form a complete or overcomplete set. 
Even if this set is not complete, but would be covering the phase space where the collective dynamics is concentrated 
one would be able to derive accurate approximate representations in this manner. There are however reasons to believe 
that the set of (generalized) Slater determinants $\Phi({\bf x}|{\bf q})$ used in the current implementation of TDGCM treatments is not sufficiently large
in the case of fission dynamics, see \cref{app:C}. 
 
The equation for the collective wave function $f({\bf q},t)$ 
is obtained from the Dirac variational principle for the wave function $\Psi( {\bf x},t)$ 
(by varying $f({\bf q},t)$ with fixed $\Phi({\bf x}|{\bf q})$)
\beq 
\delta  \int _{t_0}^{t_1} \!\!\! dt \int d{\bf x} \; \Psi^*({\bf x}, t) \left [ i \hbar \frac{\partial}{\partial t} -H\right ] \Psi({\bf x}, t)=0.
\eeq
Under the Gaussian overlap approximation
the emerging integral equation for a related to $f({\bf q},t)$ collective wave function $g({\bf q},t)$  
is transformed into a partial differential equation~\cite{Ring:2004,Pomorski:2012,Schunck:2016,Regnier:2016,Schunck:2019}  
similar to the  equation of the 
Bohr-Mottelson Hamiltonian in the space of the collective degrees of freedom ${\bf q}$
\beq
 i\hbar\frac{\partial g({\bf q},t) }{\partial t} =\left [ -\frac{1}{\sqrt{\gamma({\bf q})}}\frac{\partial}{\partial q_k}
  \frac{  \sqrt{\gamma({\bf q})}  \hbar^2 }{ 2{\cal M}_{kl}({\bf q}) }\frac{\partial}{\partial q_l} +{\cal U}({\bf q})\right ]g({\bf q},t), \label{eq:BM}
\eeq
with a collective inertia tensor ${\cal M}({\bf q})$ and collective potential energy surface ${\cal U}({\bf q})$. 
Here $\gamma({\bf q})$ is the determinant of the metric in the collective space and 
we adopted the Einstein convention for summation over repeated indices. 
The collective potential energy is obtained by minimizing the 
energy for fixed values of the collective variables ${\bf q}$
\beq
{\cal U}({\bf q})=  \langle \Phi_\text{min} ({\bf q}) | H |\Phi_\text{min} ({\bf q}) \rangle -\varepsilon_0({\bf q}),  \label{eq:U}
\eeq
where $\Phi_\text{min}({\bf x}|{\bf q})$ is obtained by minimizing the functional ${\cal V}({\bf q})$ with constraints 
and
\beq
{\cal V}({\bf q})= \textrm{min}_{\bf x} \left \langle \Phi_\text{min} ({\bf q}) \left | H 
-\sum_k\lambda_kQ_k \right |\Phi_\text{min} ({\bf q}) \right \rangle ,
\eeq
where $Q_k$ are various contraints
and the quantum average is over the intrinsic DoF ${\bf x}$.
In the above formula for the collective  potential energy the last term is due to 
the zero-point energy fluctuations, which has to be included in actual calculations to avoid double counting.
The total energy of the system is a sum of the collective kinetic energy and of the 
collective potential energy.  Collective kinetic energy is due to the presence of collective 
flow in the dynamics but it also has a contribution due to the presence of zero-point fluctuations.
The collective potential energy depends only on the spatial matter distribution, but not on any collective currents.
TDGCM or ATDHF microscopic approaches thus invoke the adiabaticity 
(no intrinsic entropy production) of the nuclear
collective shape evolution, leading to no
irreversible energy transfer from the small number of collective
DoF to the large number of intrinsic DoF.  The intrinsic system is always 
at zero temperature and entropy and all the kinetic energy is due to  the collective DoF only. 
While evolving in the collective space the nucleus is thus intrinsically 
always at zero temperature, as the local Fermi momentum distributions correspond 
to zero local temperature. 
If an energy transfer between collective 
${\bf q}$ and intrinsic ${\bf x}$ DoF would be allowed  then at given 
values of the collective variables ${\bf q}$ reached during the actual dynamics 
\beq 
 \langle \Phi({\bf q})  | H  |\Phi ({\bf q}) \rangle  >   \langle \Phi_\text{min} ({\bf q}) | H |\Phi _\text{min}({\bf q}) \rangle
\eeq
and strictly speaking in such a case the collective motion is not anymore conservative 
and a collective potential energy does not exist.  
In \cref{sec:V} and in \cref{app:C} we will expand on these aspects.

It is also natural to assimilate the collective dynamics with that of Brownian 
motion of the collective DoF in the bath of the intrinsic DoF~\cite{Frobrich:1998} 
and  describe the collective dynamics either with a classical Fokker-Planck equation or with 
an equivalent classical Langevin equation.  
Phenomenological classical Langevin
description in nuclear physics is restricted in practice to a
small space of collective variables ($\leq 5$; e.g. elongation, mass
asymmetry, neck size, and the two quadrupole deformations of the
fragments)~\cite{Moller:2001} and the number and the character of 
these collective DoF are not universally agreed upon.  Such models
require the evaluation of a potential energy surface, of a collective inertia
tensor (in most approaches), of a dissipation tensor (for Langevin
dynamics), and of a phenomenological temperature.  
When a collective Schr\"odinger equation is derived within
the TDGCM or ATDHF for a subset of collective variables the class of quantum
fluctuations generated is different from the thermal
fluctuations generated in Langevin
approaches~\cite{Frobrich:1998,Randrup:2011,Randrup:2011a,
Sierk:2017,Ishizuka:2017,Sadhukhan:2016,Sadhukhan:2017}.
However, TDGCM, ATDHF and most incarnations of the Langevin approach
rely on the assumption that the shape evolution is mostly collective
in nature and driven both by the potential energy surface and the
inertia tensor.  We provide here evidence that this assumption is invalid!

In experiments, the observed final state of nuclear fission
corresponds to a wide FFs distribution of varying charges and masses,
and a wide distribution of their kinetic and excitation
energies, angular momenta,
and parities. Various Langevin implementations and TDGCM approaches suggest that the fluctuations
around the most probable trajectory determine the distributions of the
FFs on mass, charge, kinetic, excitation energies.  
Accounting for fluctuations is also important quantum mechanically 
for totally different and unrelated reasons, 
to restore spontaneously broken symmetries. As Langevin-type
simulations and the path-integral formulation both demonstrate, the
presence of fluctuations along the entire trajectory is crucial, and
the presence of initial state fluctuations alone
is of little consequence.  We will demonstrate that since the nuclear
collective dynamics from saddle-to-scission is similar to that of a very viscous
fluid, the role of fluctuations only at the start of trajectories is
quickly erased, thus in total disagreement with the results of Ref.~\cite{Tanimura:2017}, see also \cref{app:E}.

In statistical scission-point models there is no dynamics, only the competition between FFs
configurations at the scission point are
considered~\cite{Wilkins:1972,Wilkins:1976,Lemaitre:2015}, a model to which our
results lend partial support.  
In a statistical scission-point model a full thermalization of the intrinsic DoF is implied. 
However, it is not obvious that all possible equilibrated configurations can be reached dynamically 
during the evolution from saddle-to-scission.
On the other hand, our results lend some theoretical support to the 
overdamped Brownian motion model of  
Randrup et al.~\cite{Randrup:2011,Randrup:2011a,Randrup:2013,Ward:2017,Randrup:2018}.

\section{Theoretical Framework} \label{sec:III}

Our theoretical framework is called the time-dependent superfluid local density 
approximation (TDSLDA), which is an extension of TDDFT  to superfluid 
systems~\cite{Bulgac:2011a,Bulgac:2013a,Bulgac:2019}. 
DFT, which formally looks similar to the Hartree approximation, is used in the Kohn-Sham 
implementation, often referred particularly in condensed matter and chemistry literature as LDA (local density approximation). 
As a natural extension of Kohn-Sham LDA
SLDA stands the superfluid LDA. SLDA equations formally appear 
as local Hartree-Fock-Bogoliubov (HFB) or Bogoliubov-de Gennes equations. 
The equations for the single-particle (sp) wave-functions are not obtained from the 
expectation of a Hamiltonian with interparticle interactions using 
(generalized) Slater determinants, but from an energy density functional, which 
in accordance with the DFT philosophy  
should include all possible correlations.

Within TDSLDA, the evolution of the quasi-particle wave functions (qpwfs)  is governed by the equations:
\begin{align} \label{eq:tdslda}
i\hbar \frac{\partial}{\partial t}
\begin{pmatrix}
u_{k\uparrow}  \\
u_{k\downarrow} \\
v_{k\uparrow} \\
v_{k\downarrow}
\end{pmatrix}
=
\begin{pmatrix}
h_{\uparrow \uparrow}  & h_{\uparrow \downarrow} & 0 & \Delta \\
h_{\downarrow \uparrow} & h_{\downarrow \downarrow} & -\Delta & 0 \\
0 & -\Delta^* &  -h^*_{\uparrow \uparrow}  & -h^*_{\uparrow \downarrow} \\
\Delta^* & 0 & -h^*_{\downarrow \uparrow} & -h^*_{\downarrow \downarrow} 
\end{pmatrix}
\begin{pmatrix}
u_{k\uparrow} \\
u_{k\downarrow} \\
v_{k\uparrow} \\
v_{k\downarrow}
\end{pmatrix},
\end{align}
where we have suppressed the spatial $\bm{r}$ and time coordinate $t$, and $k$  
labels the qpwfs (including the isospin) $[u_{k\sigma}(\bm{r},t), v_{k\sigma}(\bm{r},t)]$, 
with $\sigma = \uparrow, \downarrow$ the z-projection of the nucleon spin. 
The sp Hamiltonian $h_{\sigma\sigma'}(\bm{r}, t)$, and the 
pairing field $\Delta(\bm{r}, t)$ are functionals of various neutron and proton 
densities, which are computed from the qpwfs, see 
Ref.~\cite{Jin:2017} for technical details. No proton-neutron pairing is assumed 
in the present study, and the pairing field is singlet in character. 
A TDSLDA extension to a more complex pairing mechanism is straightforward.

\begin{figure}
\includegraphics[clip,width=1\columnwidth]{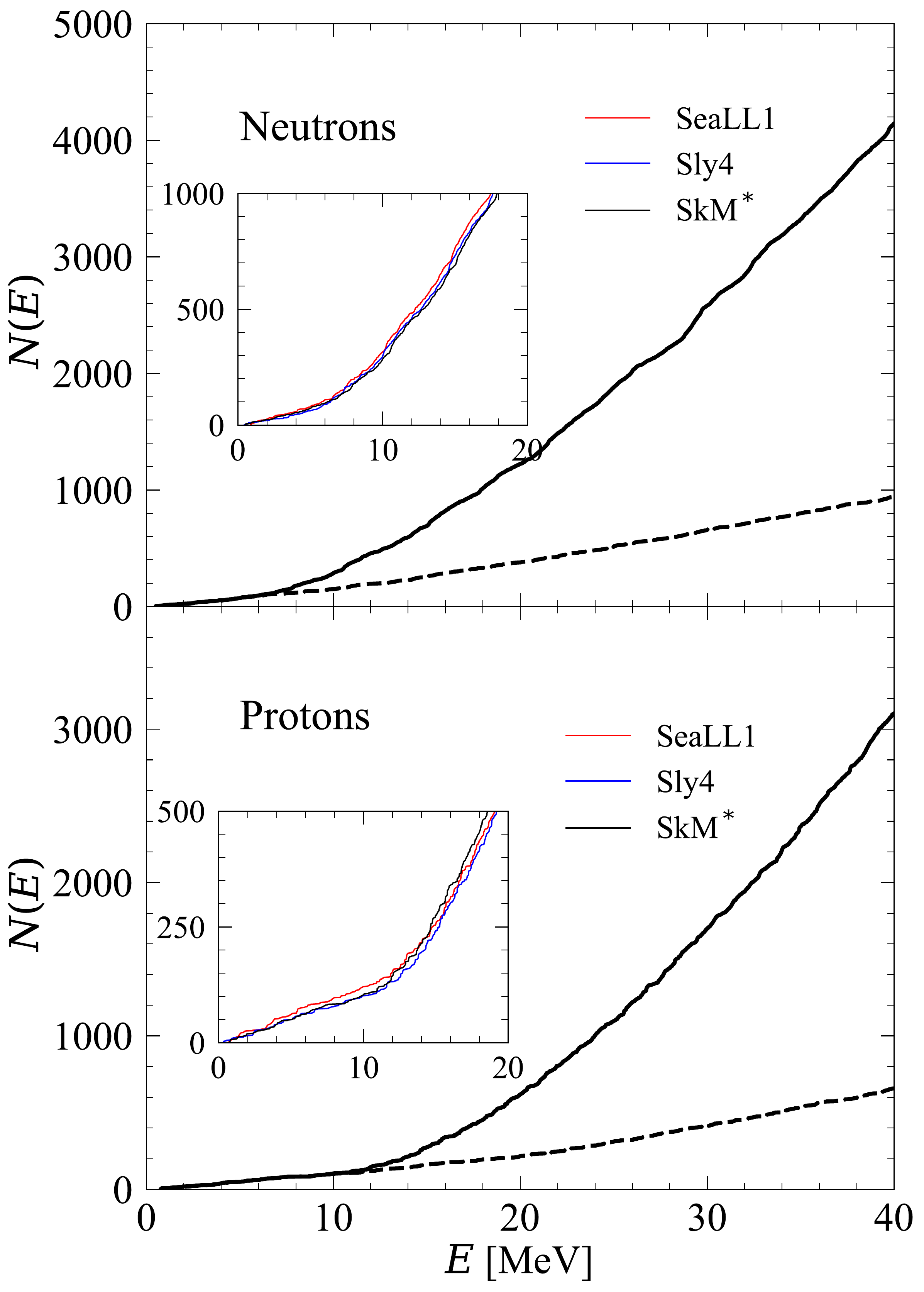}
\caption{\label{fig:Eq} (Color online) 
The cumulative number of quasiparticle states for neutrons and protons obtained in a full diagonalization
of the (initial) stationary quasiparticle Hamiltonian with constraints, see \cref{eq:eqp_static}, for neutrons and protons in a 
discrete variable representation used in (TD)SLDA  (solid line) and 
using the configuration space approach~\cite{Perez:2017} (dashed line) in case of  SkM* NEDF.
The insets  show the cumulative number of quasiparticle energies 
for SeaLL1(red), SLy4 (blue), and SkM* (black) NEDFs respectively.
The maximum quasiparticle energy is $\approx 400$ MeV 
in (TD)SLDA while in HFBTHO is $\approx 100$ MeV. The total number of 
quasiparticle states, for either neutrons or protons,
is $2N_xN_yN_z=2\times24^2\times 48=$ 55,296 in the present 
numerical implementation. In SLDA the single particle level density 
increases sharply above the nucleon separation energy, as physically expected. 
As a result, above the nucleon separation energy the configuration space approach severely underestimates
the single particle level density. }
\end{figure}

\begin{figure}
\includegraphics[clip, width=.8\columnwidth]{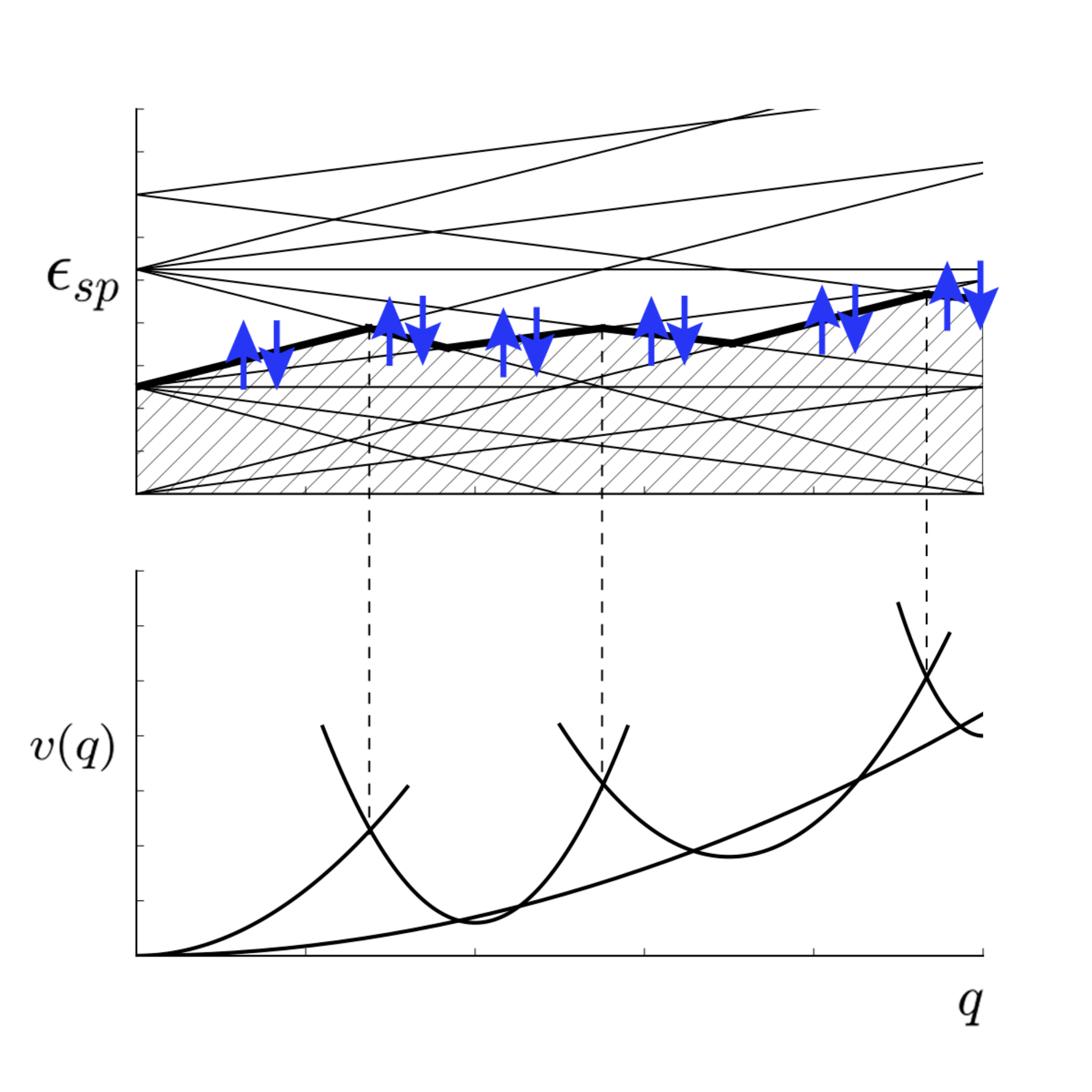}
\caption{\label{fig:esp_q} 
(Color online) Schematic evolution of sp levels of nucleons 
(upper panel) and the total nuclear energy (lower panel)  as a function of 
deformation parameter $q$ \cite{Bertsch:1980,Barranco:1990}. The thick line 
represents the Fermi level and the up/down arrows depict the Cooper pairs of 
nucleons on the Fermi level only, in time-reversed orbits $(m, -m)$. 
}
\end{figure}

A definite advantage of the TDSLDA approach is the size of the quasi-particle space. In any numerical
solution of quantum mechanical equations the relevant question is how many ``basis states'' should 
one include in the analysis in order to ensure a physically correct description of the dynamics. In 
dynamical study of fission one places the nucleus on a spatial rectangular lattice. 
One needs a simulation box spatially large enough to accommodate both 
the mother and the receding FFs until they do not significantly influence each other. At the same time
the single particle momenta allowed should be high enough to faithfully describe the 
single particle dynamics. The number of needed ``basis states'' can then estimated from a simple
formula, see Ref.~\cite{Bulgac:2013} and also \cref{app:A} for further details
\bea
&& {\cal N}_\text{sp}= 2\times \frac{ L_xL_yL_z \times (2p_c)^3}{(2\pi\hbar)^3}= 2N_xN_yN_z,\\
&&L_{x,y,z}=N_{x,y,z}l, \quad p_c =\frac{\pi\hbar}{l},
\eea
 where the factor  2 accounts for the spin, 
 $L_{x,y,z}=N_{x,y,z}l$ are the side lengths of the spatial simulation box, $N_{x,y,z}$ are the number 
 of lattice points in each spatial direction, $l$ is the lattice constant, and $p_c$ is the single particle momentum cutoff.
 In \cref{fig:Eq} we illustrate the difference between the size of the Hilbert space in a dynamical calculation and the
 size of the Hilbert space used in a configuration space approach~\cite{Perez:2017} of stationary states. 
 (The size of the quasiparticle
 Hamiltonian for either neutrons or protons is $4N_xN_yN_z$ when placed on a spatial lattice.)
 Most of the quasiparticle states are initially unoccupied, but 
 during the dynamics singe particle levels move up and down and mix,
 see \cref{fig:esp_q}, and initially unoccupied 
 high lying states are occupied. 
 
In our proof-of-concept study~\cite{Bulgac:2016} we chose rather
arbitrarily to use the SLy4 nuclear energy density functional (NEDF)
\cite{Chabanat:1998}, which accurately describes a large body of
nuclear observables throughout the nuclear mass table, even though
this functional is not particularly popular among fission
practitioners.  So far there is no deep understanding of why the
properties of various NEDFs used for fission calculations are 
responsible for the agreement or the disagreement with observations.
However, as Meitner and Frisch~\cite{Meitner:1939} and subsequent studies have shown only a 
small number of basic nuclear  properties 
(nuclear incompressibility, surface tension, and Coulomb 
interaction) were need to understand the qualitative and to some extent the quantitative
features (notably, the energy released in fission) of nuclear fission. Accounting for spin-orbit interaction and pairing 
correlations~\cite{Strutinsky:1967,BRACK:1972,Bjornholm:1980} was sufficient 
to further explain many of the remaining properties of the fission dynamics, 
e.g. asymmetric fission, odd-even staggering effects, etc. Many 
other details of the several hundreds of existing NEDFs differ often greatly, 
but never lead to significant improvements over treatments based only on describing 
the basic nuclear properties enumerated above~\cite{Moller:1995,Moller:2001,Moller:2004,Moller:2009,Moller:2012}.

\begin{figure}
\includegraphics[clip,width=\columnwidth]{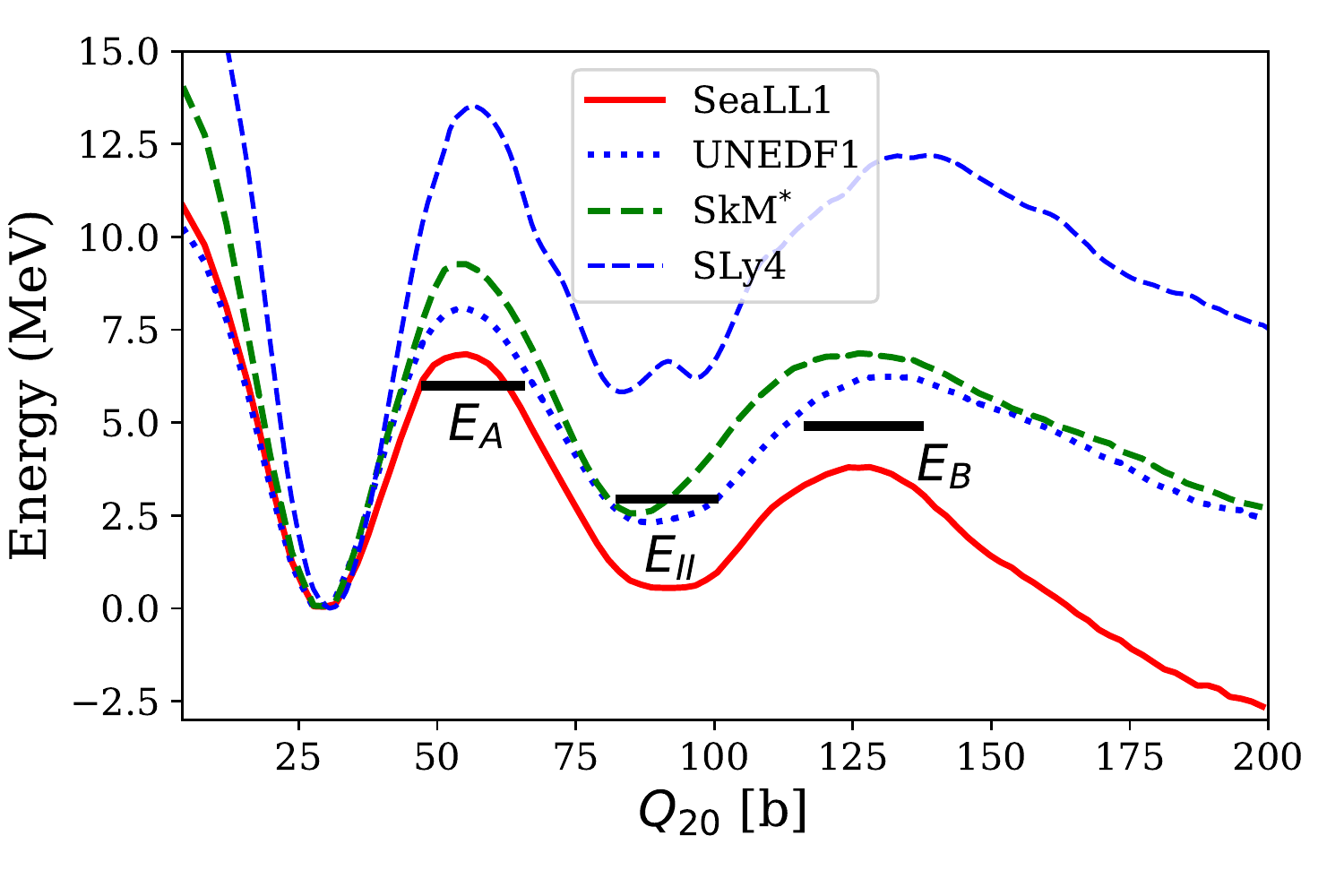}
\caption{\label{fig:paths} (Color online) 
 Fission pathway for \ce{^{240}Pu} along the mass
    quadrupole moment $Q_{20}$ calculated with
    \gls{SeaLL1}, SkM*, and UNEDF1. With black horizontal lines  labeled by 
    $E_A$, $E_B$, and $E_{II}$ we show the values of the inner and outer 
    fission barriers and the energy of the fission isomer with respect to the ground state energy. }
\end{figure}

Hundreds of NEDFs have been introduced~\cite{Dutra:2012}, depending on
a large number of parameters, and the fitting criteria used are not
universally established.  SkM*~\cite{Bartel:1982} and 
UNEDF1~\cite{Kortelainen:2012} NEDFs, have been designed to
accurately describe fission properties and the profiles of the corresponding 
potential energy surface are very similar.  We have adopted SkM* in this
study. The other NEDF we chose is the recently developed
SeaLL1~\cite{Shi:2018}, which unlike the hundreds of NEDFs introduced
in the literature, relies on the smallest number of fitting
parameters, all of them tied to basic nuclear properties, and delivers
one of the best global descriptions of a large number of nuclear
properties (masses, charge radii, compressibility, surface tension, isospin symmetry,
shell structure, pairing, two-nucleon separation energies, etc.).  In \cref{fig:paths} we compare the profiles
of the fission pathways for SLy4, SeaLL1, UNEDF1, and SkM*.
These calculations were performed with the HFBTHO
\gls{DFT} solver~\cite{Perez:2017}, triaxiality is not included and the height of the
first fission barrier is typically overestimated for these functionals
by about \SI{2}{MeV} or even more in case of SLy4~\cite{Schunck:2014}.  
Compared with SkM* and
UNEDF1, SeaLL1 underestimates the excitation energy of the fission
isomer ($E_{I\!I} = \SI{0.54}{MeV}$ compared with an experimentally extracted value
of 2.8 MeV) and the heights of the extracted fission barriers
($E_{A} = \SI{6.84}{MeV}$ vs.\@ \SI{6.05}{MeV}, and
$E_B = \SI{4.20}{MeV}$ vs.\@ \SI{5.15}{MeV}, respectively, for the
inner and outer barriers) agree within \SI{1}{MeV}.

Both SkM* and
UNEDF1 were constrained specifically on the height of the first
fission barrier (SkM*) or excitation energy of the fission isomer
(UNEDF1), while no specific information
for nuclei at large deformation in constructing SeaLL1 was used.  
Without any such constraint, the
resulting {NEDF} is still in reasonable agreement with experimental results,
especially the height of the two barriers.  Our results are definitely
better than predictions with SLy4 \cite{Chabanat:1998}, which
predicts the second fission barrier higher than the first one, 
after one includes beyond mean field corrections~\cite{Bender:2004}.  
The differences between the extracted ands calculated  fission barriers with 
NEDF designed for fission can reach \SI{2.5}{MeV}, see Ref.~\cite{Kortelainen:2014_2}, where fission barriers and the
energy of the second isomer in chains of \ce{Ra}, \ce{Th}, \ce{U},
\ce{Pu}, \ce{Cm}, and \ce{Cf}, are compared to the UNEDF1-2, Gogny
D1S~\cite{Berger:1989}, and FRLDM~\cite{Moller:2009} functionals.   
In a recent study of the surface energy coefficient
for 76 parameterizations of the Skyrme
{NEDF}~\cite{Jodon:2016} it was shown that the energy of the fission
isomer and the height of the outer fission barrier vary
by several \si{MeV}s with respect to the ground state energy. 

Apart from exploring the sensitivity of fission dynamics characteristics on the
NEDF properties, it is also imperative to study the sensitivity of TDDFT trajectories
with respect to their initial conditions. Bohr's compound nucleus model~\cite{Bohr:1936} 
would suggest that initial conditions in general should not matter. Initial conditions near
the outer fission barrier might however matter, as the dynamics from the outer fission barrier onward is 
faster than starting from the ground state configuration after capturing a neutron.
On the other hand Feyman's path-integral approach and 
the phenomenological Langevin approach, see \cref{sec:I}, 
would suggest that fluctuations along the fission path, not initial fluctuations should dominate the dynamics. 
Recently a claim was made
that fluctuations in an ensemble of peculiarly chosen initial conditions alone with 
absolutely no fluctuations along the fission path would be
sufficient in order to describe the FFs yields and the total kinetic 
energy (TKE) distributions~\cite{Tanimura:2017}, a claim which our results 
conspicuously do not support. 

\begin{figure*}
\includegraphics[width=1.\columnwidth]{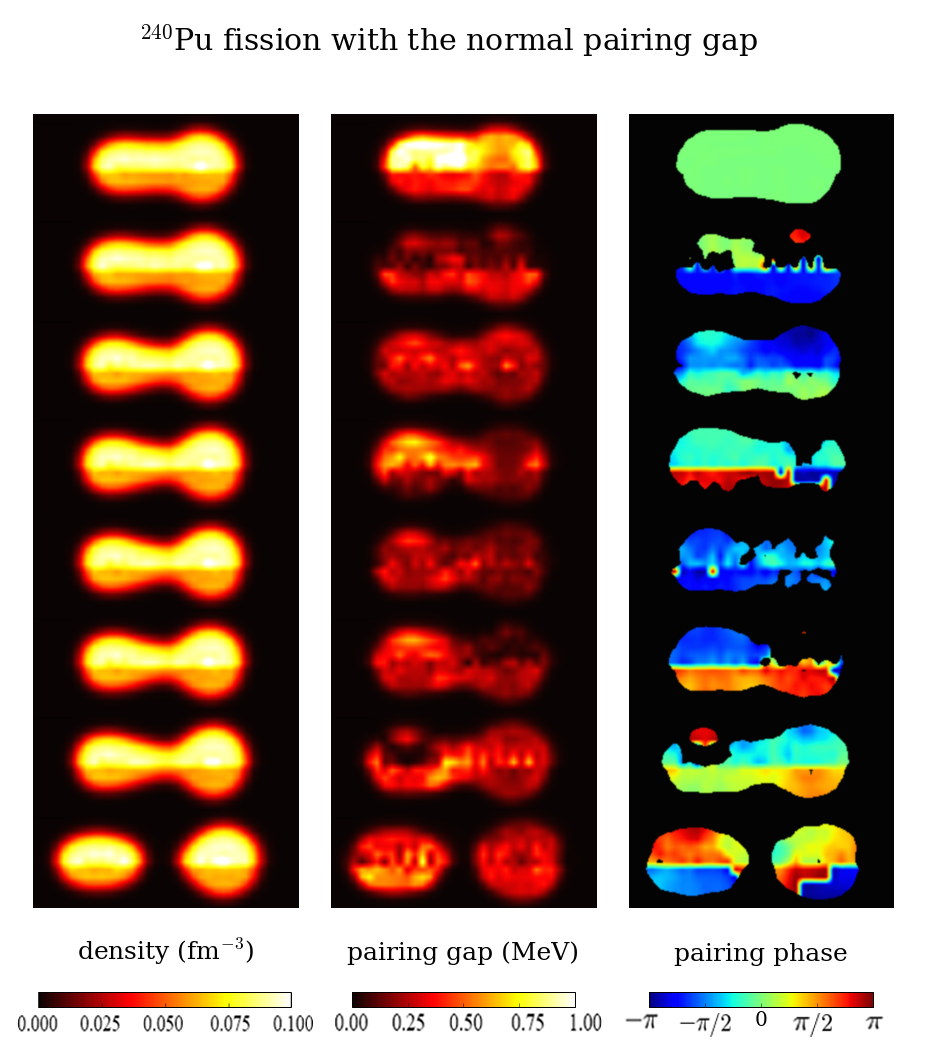}
\includegraphics[width=1.\columnwidth]{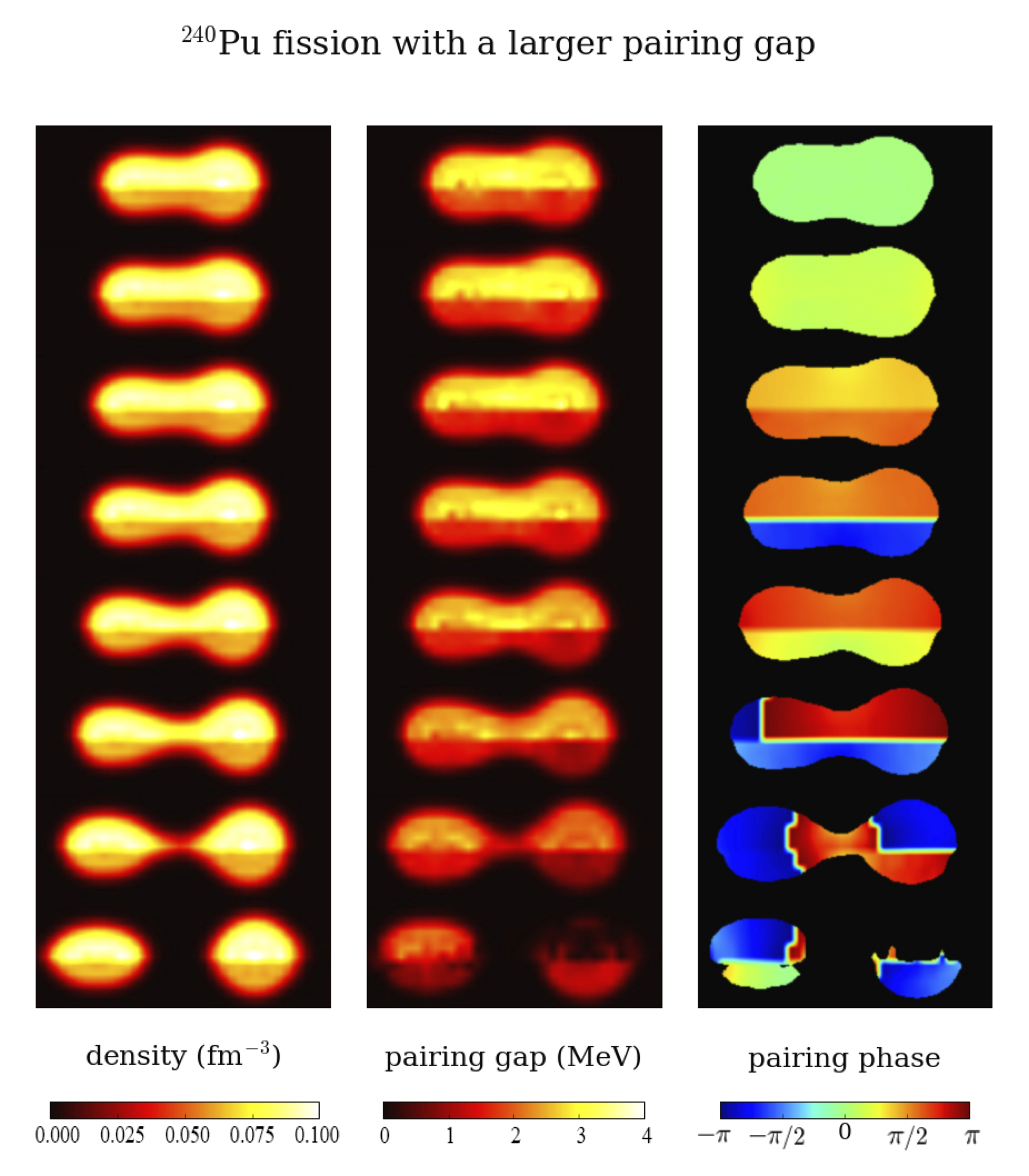}
\caption{\label{fig:pairing}
(Color online) The left three columns shows the induced fission of $^{240}$Pu 
with normal pairing strength, which lasts up to 14,000 fm/c($\approx 47 \times 
10^{-21}$ s) from saddle-to-scission. The columns show sequential frames of the 
density (1st column), the magnitude of the pairing field (2nd column), and the 
phase of the corresponding pairing field (3rd column). The upper/lower part of each frame 
shows the neutron/proton density, the magnitude of neutron/proton pairing 
fields, and of the phase of the pairing field respectively \cite{Bulgac:2016}.
The right three columns shows the corresponding snapshots of the induced 
fission of $^{240}$Pu with enhanced pairing strength, which lasts about 1,400 
fm/c.
}
\end{figure*}

\section{Role of pairing correlations} \label{sec:pairing}

The essential role of pairing correlations in nuclear shape dynamics has
been addressed qualitatively in the past.  A simplified picture was
presented by Hill and Wheeler~\cite{Hill:1953} and was later refined
by Bertsch~\cite{Bertsch:1980,Barranco:1990,Bertsch:1997,Bertsch:2018}, who
emphasized the crucial role played by the pairing interaction.  While a
nucleus deforms, the sp levels move up and down, and
typically cross in the absence of pairing, as shown in \cref{fig:esp_q}.  The sp
occupation probabilities remain unchanged if levels cross, which in case of large
prolate shapes leads to a very oblate Fermi surface and thus to a volume energy excitation
of the nucleus.  As Meitner and Frisch~\cite{Meitner:1939} have correctly assumed, 
during fission the nuclear volume practically does not change, only the surface area increases.
Thus a volume type of energy excitation is excluded.
As sp levels are doubly occupied due to Kramers
degeneracies, only the pairing short-range interaction can provide a
very effective mechanism to move simultaneously a pair of nucleons in time-reversed
states from one level to another at a (avoided)
crossing~\cite{Bertsch:1980,Barranco:1990,Bertsch:1997,Bertsch:2018}.  The
probability of such transitions is particularly enhanced in the presence of a
Bose-Einstein condensate of Cooper pairs, but such transitions remain important even in 
the absence of the condensate.

Apart from the arguments that the nuclear volume does not change and therefore the
local Fermi sphere should remain spherical, the fact that fission is hindered in the 
absence of pairing correlations (at least at the mean field level) was demonstrated 
recently by \textcite{Tanimura:2015,Goddard:2015,Goddard:2016}.

To illustrate the crucial role played by 
the pairing correlations in fission dynamics, we performed a TDSLDA simulation 
with an initial configuration identical to the S3 case of 
Ref.~\cite{Bulgac:2016}, but enforcing stronger pairing correlations by 
increasing the absolute value of bare coupling constant $g_0$. The 
corresponding average neutron and proton pairing gaps in the initial state 
increase from 0.73 and 0.33 MeV to 2.57 and 1.62 MeV respectively. By increasing the 
strength of the pairing field, the fission dynamics proceeds approximately 
10x faster. \cref{fig:pairing} shows the snapshots of the number density, 
magnitude of pairing field, and phase of pairing field for neutron and proton 
respectively in these two simulations. The left three columns in 
\cref{fig:pairing} show the induced fission of $^{240}$Pu with 
realistic pairing strength, which lasts up to 14,000 fm/c from 
saddle-to-scission, while the right three columns 
show the dynamics with an enhanced pairing strength, which lasts only about 
1,400 fm/c. In the case with normal pairing strength, the pairing field on the 
way from saddle-to-scission fluctuates noticeably in magnitude and phase. 
Therefore, strictly speaking the pairing field during its time evolution stops 
being a superfluid condensate of Cooper pairs, which otherwise would exhibit a 
long-range-order. However, in the case with larger pairing strength, the 
pairing field shows the expected characteristics of a slowly evolving 
superfluid condensate, the nuclear fluid behaving almost like a perfect or 
ideal fluid. This pattern was also observed in case of collision of two 
superfluid heavy-ions~\cite{Bulgac:2017,Scamps:2019}.  
Even though realistic pairing correlations are relatively weak, they still provide 
the essential ``lubricant'' for the saddle-to-scission evolution to take place.

\section{Results} \label{sec:V}

\begin{figure}[!hbp]
\includegraphics[width=1.1\columnwidth]{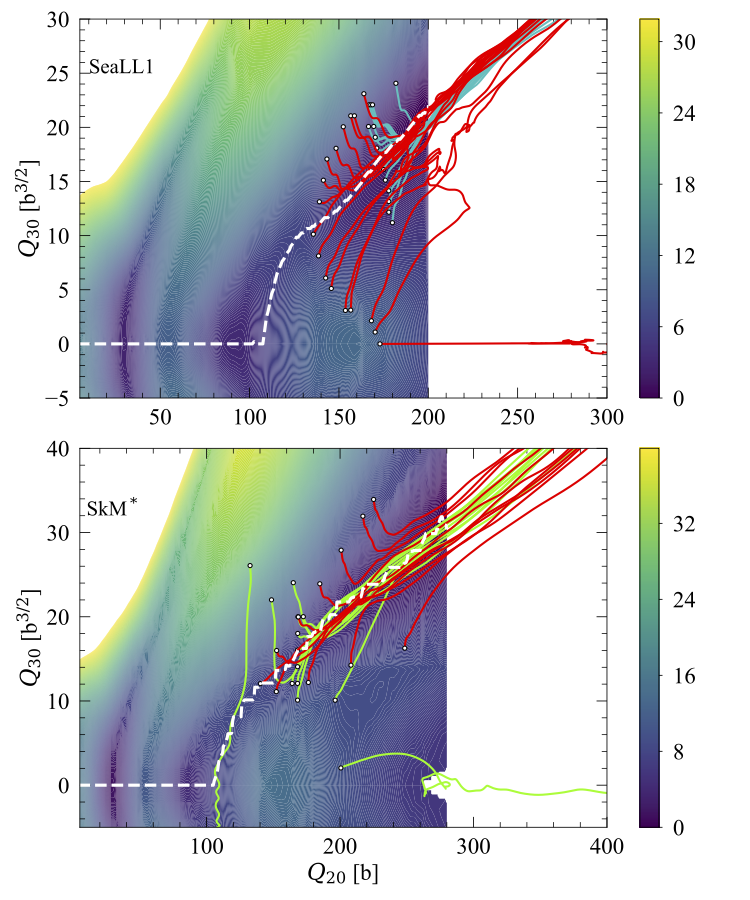}
\caption{\label{fig:pes}
(Color online) Fission trajectories for SeaLL1 (upper panel) and SkM* (lower pannel).
The red (SeaLL1-1) and cyan lines (SeaLL1-2)  correspond to initial 
configurations slightly above to and lower than the outer fission barrier.  
The SeaLL1-1sy trajectory had a small initial left-right asymmetry.  The green lines in the lower panel 
correspond to SKM$^*$-1asy and and the red lines to the trajectories obtained with SkM$^*$-2asy.
The  white dashed lines show the path linking the ground state minimum, 
the fission isomer minimum, and the inner and the outer saddle points. In both panels 
we also show an example of a symmetric fission trajectory, and in the low panel also 
of a trajectory which wondered inside the fission isomer well.
}
\end{figure}

We have chosen an ensemble of initial conditions in the
$Q_{20},Q_{30}$ collective coordinates, in total 60 different initial
conditions (including the four trajectories from Ref.~\cite{Bulgac:2016}), 
as shown in \cref{fig:pes}. These initial conditions are prepared by  
constrained HFB calculations with the HFBTHO solver~\cite{Perez:2017}.
Using these densities we generated 
the raw qpwfs used as initial conditions in the time-dependent 
simulations, in the absence of any constraints, see \cref{app:A} for details.

One set of initial conditions (SeaLL1-1) corresponds to
configurations of $^{240}$Pu with mean excitation energy/variance
 $7.9/1.7$ MeV in the neighborhood of the outer
saddle point, which can be reached in low energy neutron induced
fission. The other set of initial conditions (SeaLL1-2) corresponds to
a mean excitation energy/variance  $2.6/1.8$ MeV, which can
be reached either in spontaneous fission or with photo-excitation
excitation of $^{240}$Pu. The third set of initial conditions (SkM*-1)
is similar to SeaLL1-1, with mean excitation energy/variance
$8.2/3.0$ MeV, but with an increased pairing strength. The fourth
set (SkM*-2) was characterized by a realistic pairing strength.
In the simulations with SLy4 NEDF~\cite{Chabanat:1998} and SkM*,  
we neglected the correction term $1/A$ for the center-of-mass kinetic energy
in the sp kinetic energy $1-1/A$.  Without this correction term 
these NEDFs satisfy local Galilean invariance.
We have checked that this term has a negligible influence 
on the profile of the potential energy surface.

\subsection{Fission fragments properties} \label{sec:V A}

The most surprising outcome of these
simulations is that in all these sets of initial conditions, which
correspond to vastly different initial values of $Q_{20},Q_{30}$, we
observed a very strong focusing effect and the final states are
remarkably similar, see \cref{fig:pes}.
The heavy fragments has neutron and proton numbers between those of  the double magic 
\ce{^{132}}{Sn} ($N=82,Z=50$) and of the octupole shaped $^{144}$Ba ($Z=56$,  
$N=88$), and has a shape quite close to spherical. 
The lighter fragment has an elongated shape (see also \cref{tab:deform} ). 
\textcite{Simenel:2018} have recently shown that the octupole shell stabilization 
of nuclei close to $^{144}$Ba with $Z=56$ drive the fission dynamics towards 
proton numbers larger than 50, as we also appear to confirm. As we show below, see \cref{sec:V D}
and \cref{fig:q2-q3}, at scission both FFs have a significant octupole deformation, 
which however relaxes after the FFs separate. 
The neutron and  proton numbers (and thus the mass) of the FFs match pretty well to the mean 
values of the experimental systematics, but show a very small dispersion, see \cref{table:ff}. 

The strong focusing effect we have establish in the present study is in stark contrast with the results of \textcite{Tanimura:2017}. 
The authors of that study generated an ensemble of initial conditions according to the stochastic 
mean field model of \textcite{Ayik:2008}. In the stochastic mean field model the nucleon single-particle wave-functions (spwfs) are 
evolved using the old-fashion TDHF method and the only difference is in considering an ensemble of 
different initial conditions for the one-body density matrix~\cite{Ayik:2008,Tanimura:2017} and \cref{app:E}, 
which result in ensemble of initial states with different initial energies and quadrupole $Q_{20}$ 
and octupole $Q_{30}$ moments. In this respect our choice of various initial conditions spread over
a significant area of the potential energy surface and the choice chosen by \textcite{Tanimura:2017} 
and the subsequent time-dependent evolution of the nucleonic spwfs are qualitatively similar, but the final results are qualitatively 
different.  We attribute these differences 
to the fact that the stochastic mean field approach leads to flagrant violations of the Pauli principle, see \cref{app:E}.

The TKE and total excitation energy (TXE) are also calculated.
The TKE at a relatively large finite separation between the fragments ($\approx 25$ fm) 
and in the center-of-mass reference frame is evaluated with the formula
\begin{subequations}\label{eq:TKE}
\begin{align}
E_\textrm{TKE} = \frac{1}{2}mA_{\mathrm{H}} \vec{v}_{\mathrm{H}}^2 + \frac{1}{2}mA_{\mathrm{L}} \vec{v}_{\mathrm{L}}^2 + E_{\mathrm{Coul}}, 
\end{align}
with the velocity of the fragment $f = H, L$ given by 
\begin{align}
\vec{v}_f = \frac{1}{mA_f} \int_{V_f} d{\bf r} \vec{j}(\vect{r}), \quad A_f = \int_{V_f} d{\bf r} n(\vect{r}) , 
\end{align}
where $\vec{j}(\vect{r})$ and $\n(\vect{r})$ are the total current
and number densities respectively, and  the integral is performed over the 
appropriate half-box $V_f$ where each fragment is located. The Coulomb interaction energy (direct term only) is given by 
\begin{align}
E_{\mathrm{Coul}} = e^2 \int_{V_{\mathrm{H}}} d{\bf r}_1\int_{V_{\mathrm{L}}} d{\bf r}_2 \frac{n_p(\vect{r}_1)n_p(\vect{r}_2)}{\lvert \vect{r}_1 - \vect{r}_2 \rvert}
\end{align}
where $n_p(\vect{r})$ is the proton number density.
\end{subequations}

The excitation energy of each FF is calculated by extracting 
the computed ground state energy of each FF from the energy 
of each FF in its rest frame. 
The FF ground state energy is computed with the HFBTHO code~\cite{Perez:2017} 
for the corresponding FF neutron and proton numbers.
The proton and neutron numbers of FFs are evaluated from 
\beq
Z_f= \int_{V_f}d{\bf r} n_p({\bf r}), \quad N_f=\int_{V_f}d{\bf r}n_n({\bf r}),
\eeq
 and the TXE is evaluated from
\begin{align}\label{eq:TXE}
E_\mathrm{TXE} = E^*_{\mathrm{H}} + E^*_{\mathrm{L}}.
\end{align}

The energy variance of the TKE and TXE 
are only slightly larger than those of the initial energies.  Compared to SeaLL1-2, 
the SeaLL1-1 starts at a larger excitation, and it has a longer average 
saddle-to-scission time ($\tau_{s\to s}$) and larger average TXE 
for the fission fragments, while their average TKEs are almost the same. 
When comparing the FFs characteristics emerging from simulations with 
SeaLL1 in \cref{fig:pairing} (see also \cref{sec:pairing}) we notice that in the case of enhanced pairing 
the scission configuration corresponds to a longer neck and thus to a lower TKE. 
This is also confirmed 
by the results obtained with NEDF SkM*-1 (enhanced pairing strength) and SkM*-2 (realistic pairing strength). 
Another particular aspect that emerges from our simulations is the character of the excitation 
energy sharing between the light and heavy fragments. In the case of SeaLL1-1 the light fragment
has a larger excitation energy than the heavy fragment, while the case of SeaLL1-2 has 
the opposite pattern. These differences lead us to conclude that
the excitation energy sharing in the cases of spontaneous fission and
induced fission are different.    
The observed wider mass yields for $^{239}$Pu(n,f) than in
$^{240}$Pu (s.f.)~\cite{Wagemans:1984,Schillebeeckx:1992} apparently point
to differences in the over the barrier and under the barrier fission dynamics too.

\begin{table*}[t]
  \begin{ruledtabular}
    \begin{tabular}{|l|l|l|l|l|l|l|l|l|l|l|l|}
\multicolumn{1}{c|}{NEDF} & \multicolumn{1}{c|}{$E^*_{\text{ini}}$} & \multicolumn{1}{c|}{$\text{TKE}$}  &\multicolumn{1}{c|}{$N_{\rm H}$} & \multicolumn{1}{c|}{$Z_{\rm H}$} & 
\multicolumn{1}{c|}{$N_{\rm L}$} & \multicolumn{1}{c|}{$Z_{\rm L}$}& \multicolumn{1}{c|}{$E^*_{\rm H}$} & \multicolumn{1}{c|}{$E^*_{\rm L}$} & \multicolumn{1}{c|}{TXE} & \multicolumn{1}{c|}{TKE+TXE} &\multicolumn{1}{c}{$\tau_{s\rightarrow s}$ (fm/c)}  \\ \hline  
SeaLL1-1asy    & $7.9(1.7)$ & $177.8(3.1)$ & $83.4(0.4)$ & $53.2(0.4)$ & $62.9(0.5)$ & $41.1(0.4)$ & $17.1(3.0)$ & $20.3(2.0)$ & $37.4(3.1)$ & $215.2(2.5)$ & $2317(781)$\\
SeaLL1-2asy   & $2.6(1.8)$ & $178.0(2.3)$ & $82.9(0.4)$ & $52.9(0.2)$ & $63.3(0.5)$ & $41.5(0.3)$ & $19.5(3.8)$ & $14.0(1.9)$ & $33.5(5.1)$ & $211.5(3.3)$ & $1460(176)$\\
SeaLL1-sy & $9.2$ &          $147.1$      & $77.5$      & $48.9$ &    $68.8$     &       $45.4$   & $45.2$      &  $29.0$      & $74.2$      & $221.3$    &  $10103$ \\ 
    
SkM$^*$-1asy& $8.2(3.0)$ & $174.5(2.5)$ & $84.1(0.9)$ & $53.0(0.5)$ & $61.8(0.9)$ & $40.9(0.5)$ & $16.6(3.1)$ & $14.9(2.3)$ & $31.5(3.8)$ & $206.0(2.4)$ & $1214(448)$\\
SkM$^*$-1sy & $9.6$      & $149.0$      & $73.4$      & $47.2$      & $72.6$      & $46.7$      & $29.4$      & $28.5$      & $57.9$      & $206.9$      & $3673$ \\
SkM$^*$-2asy& $8.1(0.2)$ & $182.8(4.4)$ & $82.6(1.0)$ & $52.4(0.6)$ & $63.6(1.0)$ & $41.7(0.5)$ & $14.3(3.9)$ & $13.0(3.0)$ & $27.3(3.4)$ & $210.1(1.8)$ & $1349(309)$\\
\end{tabular}
    \end{ruledtabular}
    \caption{\label{table:ff}
The NEDF, the initial excitation energy $E^*_{\text{ini}}$, TKE, neutron, proton number, and excitation energies  
of the heavy and light fragments, total excitation energy of fragments TXE,  and the sum of TKE and TXE, and the average 
 saddle-to-scission times and their corresponding  variances in parentheses. All energies are in MeV and S***sy, S***asy stand for symmetric and 
 antisymmetric  channels. Using Wahl's charge systematics~\cite{Wahl:2002} and data from Ref.~\cite{Vogt:2009} one obtains for 
 neutrons $N_{L}^{syst}\approx 61$ and  $N_H^{syst}\approx 85$  and for protons $Z_{L}^{syst}\approx 40$   and $Z_H^{syst}\approx 54$, and 
TKE$^{syst}=177 \ldots 178$ MeV from Ref.~\cite{Madland:2006}.   } 
    \end{table*}

\begin{table*}[!htb]
\begin{tabular}{l|r|r|r|r|r|r|r|r|r|r|r}
\hline \hline
NEDF  & $T_{\rm L}$ [MeV] & $T_{\rm H}$ [MeV] & $T_{\rm L}$ [MeV] & $T_{\rm H}$ [MeV] & 
     $Q^{\rm L}_{20}$ [b] & $Q^{\rm H}_{20} $ [b] & $Q^{\rm L}_{30}$ [b$^{3/2}$] & $Q^{\rm H}_{30}$ [b$^{3/2}$] & 
     $(c/a)_H$ & $(c/a)_L$ & $\tau_{s\rightarrow s}$ [fm/c] \\ \hline 
SeaLL1-1 & $ 1.40(0.07)$ & $1.11(0.08)$ & $1.28(0.07)$ & $ 1.16(0.07)$ & 
           $15.7(0.9)$   & $2.6(0.5)$   & $0.08(0.17)$ & $-0.20(0.06)$ & 1.06(0.01) & 1.59(0.03)&  $2392(800)$\\
SeaLL1-2 & $ 1.15(0.08)$ & $1.19(0.12)$ & $1.00(0.08)$ & $1.21(0.08)$  &  
           $17.1(1.1)$   & $2.6(0.6)$   & $0.23(0.08)$ & $-0.19(0.06)$ & 1.06(0.01) & 1.63(0.03) & $1460(176)$\\
SeaLL1-sy & $1.54$      &  $1.99$        &  &  &   
            27.4         &       27.0   &    0.9       & -1.1          & 1.87 & 1.73 &$10103$\\                        
SkM*-1asy& $ 1.20(0.09)$ & $1.10(0.10)$ &  &  &  
           $11.3(1.3)$   & $3.5(0.9)$   & $0.1(0.1)$   & $-0.4(0.1)$   & 1.08(0.02) & 1.42(0.04)& $1214(448)$\\
SkM*-1sy &  1.56        &        1.55  &  &  &   
            24.2         &       25.6   &    0.9       & -1.0          & 1.72 & 1.75 &$3673$\\ 
SkM*-2asy& $ 1.11(0.14)$ & $1.02(0.14)$ &  &  &  
           $14.5(1.7)$   & $2.3(0.7)$   & $0.09(0.08)$   & $-0.3(0.1)$   & 1.05(0.02) & 1.53(0.06)& $1349(309)$\\
           
\hline \hline
\end{tabular}
\caption{\label{tab:deform} Internal temperatures for the light $T_{\rm L}$ and 
heavy $T_{\rm H}$ fragments computed according to the simple estimate (columns 
2 and 3) or finite-temperature HFB calculations (columns 4 and 5). The axial 
quadrupole and octupole moments of the fragments, the ratios of the long to the short semi-axes, as well as the average 
scission times are also listed}
\end{table*}

It is instructive to 
express excitation energy of the FFs in terms of an internal temperature.
We have used two different methods to extract this temperature. In the first approach, 
we have estimated the temperatures of the light and 
heavy fragments by the simple formula $E^*_f = \tfrac{A_fT^2_f}{ a}$, where  $T_f$ 
is its temperature and $a \approx 10$~\cite{Bohr:1969}. 
Such simple estimates are often 
used in simulations of the decay of the fission fragments using either 
Hauser-Fesbach or statistical evaporation models \cite{Talou:2011,Becker:2013,
verbeke2015,verbeke2018}. 

In the second approach used to determine the FFs temperatures  we have performed full 
finite-temperature HFB calculations with the HFBTHO solver. Calculations were 
performed by constraining $N_f$, $Z_f$, $\langle\hat{Q}_{20}^f\rangle$ and 
$\langle\hat{Q}_{30}^f\rangle$ to the values extracted in the relaxed fragments,
see \cref{table:ff}, \cref{tab:deform}, and \cref{sec:V D}. 
For each individual FF, we extract the temperature from the function 
$E^{*}_f(T)$ and find the corresponding $T_f$ for the given $E^*_f$. 
This calculation is more realistic than the simple estimate, even 
though (i) by constraining only $\langle\hat{Q}_{20}\rangle$ and 
$\langle\hat{Q}_{30}\rangle$, we do not obtain exactly the same shape as the 
actual FFs and (ii) the temperature thus obtained should be 
thought of as the maximum allowable value; see discussion in \cite{schunck2015b}.
In \cref{tab:deform}, column 2 and 3 list the 
average and variance over various trajectories of the temperature of light and heavy fragments in the first approach,
and column 4 and 5 list the values in the second approach. These two approaches give comparable results
and while in SeaLL1-1 the light fragments have higher temperature than the heavy fragments in SeaLL1-2 
the opposite is true. The relaxed values of the average and of the variance 
over the ensemble of trajectories of the quadrupole and of the 
octupole moments of fragments, see \cref{fig:q2-q3} and \cref{sec:V D},   are listed in column 6 to 9. In 
column 10 and 11 the ratio of the long to the short semi-axes
of relaxed FFs are listed. 
As the initial state excitation energy increases, from SeaLL1-2 to 
SeaLL1-1, one notices that the extra energy mostly goes to the light FF.  
This suggests that the average neutron multiplicity
spectrum of spontaneous and neutron-induced fission could be noticeably 
different.

The total energy released $Q=E_\textrm{TKE}+E_\textrm{TXE}$ can be estimated alternatively with
known FFs neutron and proton numbers, c.f \cref{table:ff}, by using a liquid drop 
model mass formula for the masses of the mother and daughter nuclei, 
including also the initial excitation energy.  Using the the liquid drop mass parameters
$a_v$=-15.47 MeV, $a_s$=16.73 MeV, $a_I$=22.87 MeV, and $a_c$=0.699 MeV 
\beq
E_{gs}=a_vA+a_sA^{2/3}+a_I\frac{(N-Z)^2}{A}+a_c\frac{Z^2}{A^{1/3}},
\eeq
obtained in Ref.~\cite{Shi:2018} by fitting 2375 measured nuclear masses~\cite{Audi:2012} with an energy rms 3.30 MeV, one obtains for 
SkM*-2asy trajectories Q=205 MeV, as compared to 210.1 MeV from the simulation, see \cref{table:ff}.
With the parameters $a_v$=-15.77 MeV, $a_s$=17.50 MeV, $a_I$=23.65 MeV, and $a_c$=0.723 MeV obtained by 
fitting the ground state masses evaluated with SeaLL1 in Ref.~\cite{Shi:2018} we obtain 
215.0 MeV for SeaLL1-1asy and 210.9 MeV for the SeaLL1-2ays, as compared to the calculated mean values
215.2 MeV and 211.5 MeV respectively, see \cref{table:ff}. This last (unoptimez) parametrization of the liquid 
drop mass formula reproduces the SeaLL1 ground state energies in mean filed 
(without beyond mean field corrections)  for 606 even-even nuclei with 
mean energy error of 0.97 MeV and an energy rms of 1.46 MeV.
 
\begin{figure}
\includegraphics[width=\columnwidth]{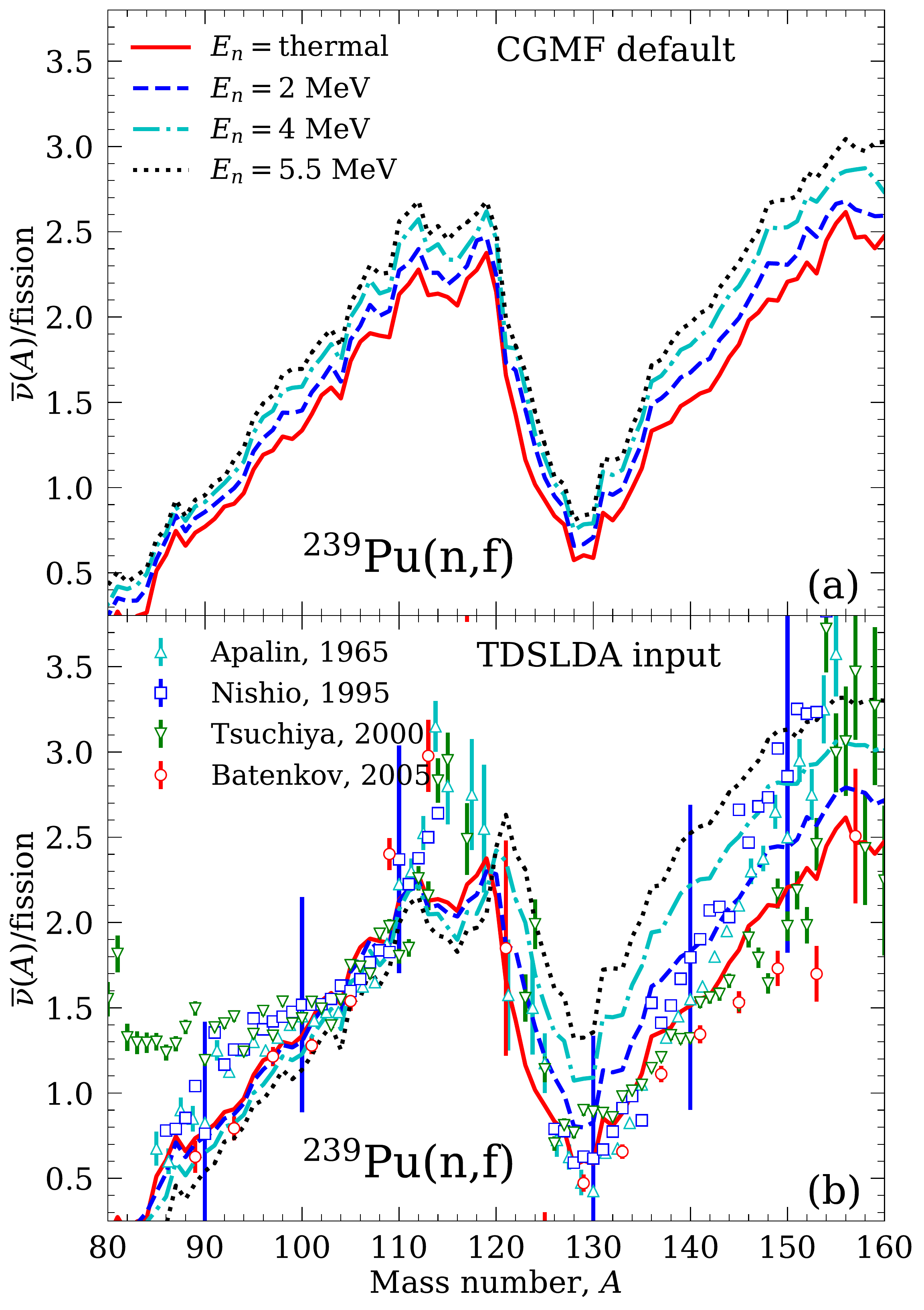}
\caption{\label{fig:nubar}
(Color online)  We compare here the average neutron multiplicity $\bar{\nu}(A)$ emitted by FFs in the case of a default 
CGMF simulation, which assumes no $E_n$ dependence for the energy sharing,  with the one extracted using the the excitation 
energy sharing between the FFs in our calculation with NEDF SeaLL1, as a function of the equivalent incident neutron energy in 
$^{239}$Pu(n,f) reaction along with available experimental data for the reaction $^{239}$Pu(n$_\text{th}$,f) from 
Refs.~\cite{Apalin:1965,Nishio:1998,Tsuchiya:2000,Batenkov:2005}. The fragment mass $A$ is before neutron emission. 
}
\end{figure}

From the ensemble of initial conditions used to generate the fission 
trajectories in case of NEDF SeaLL1, which have a 
significant spread in initial energies  and the corresponding FFs 
excitation energies we determined the average ratio of the temperatures
$R_T=T_L/T_H$ as a function of the equivalent incident neutron energy $E_n$ in the reaction $^{239}$Pu(n,f), 
using for each FF the simple estimate $E^*_f = \tfrac{A_fT^2_f}{ a}$. 
We parameterize the energy dependence obtained in the calculations, by approximating $R_T^2$ with a linear function of $E_n$ 
(while we do not expect an overall linear dependence, this should be a good approximation for up to about 5.5 MeV incident neutron 
energies). This linear dependence is added as a multiplicative factor, $R_T(A,E)\approx R_T(A) f(E_n)$, to the 
parameterization of the energy sharing in the Cascade Gamma Multiplicity for Fission (CGMF) code, which is used to model the 
neutron and gamma emission~\cite{Talou:2011,Becker:2013,Stetcu:2016}. Experimental evidence for the energy sharing is indirectly provided 
by detecting the average number of fission neutrons emitted as a function of fragment mass. In CGMF, the default calculation 
parameterizes this sharing by assuming a mass dependence of $R_T(A)$ which is adjusted to available data (select spontaneous 
and thermal neutron induced fission reactions). The underlying assumption is consistent with 
no energy dependence for the CGFM  default parameterization, and, 
thus, when CGMF calculations are performed, the average multiplicity of neutrons emitted as a function fission fragment 
mass increases almost uniformly with increasing the incident neutron energy for both light and heavy fragments, as illustrated in 
\cref{fig:nubar}(a). However, experimental evidence in U and Np neutron induced reactions has shown that only for the heavy 
fragment the number of neutrons increases~\cite{Muller:1984,Naqvi:1986}, while the number of neutrons emitted from the 
light fragments remains constant, within experimental uncertainties. Adding the parameterization of the energy dependence from out 
TDSLDA calculations, \cref{fig:nubar}(b) shows that we obtain indeed a very similar trend with what is expected from the 
experimental data, which suggests that the TDSLDA modeling of the excitation energy sharing between fission fragments is 
reasonable.

\subsection{Collective flow energy} \label{sec:V B}

Certain rather crucial aspects of the nuclear collective motion were never elucidated
in a microscopic calculation, and were treated only
phenomenologically.  Is the character of the evolution from
saddle-to-scission adiabatic? If not, is it controlled by the one-body
and/or the two-body dissipation, and if so, to what extent?  In
phenomenological studies the strength of the one-body dissipation is
often artificially reduced and a contribution arising from the
two-body dissipation mechanism is included~\cite{Sierk:2017}.  

It is important to remember that the effect of two-body collisions is encoded in the collision integral of
the Boltzmann-Uehling-Uhlenbeck
equation~\cite{Uehling:1933,Bertsch:1988}.  On the other hand, the evolution equation for
the local number density $n({\bf r},t)$, i.e. the continuity equation
$m\dot{n}({\bf r},t)+\text{div}\,{\bf p}({\bf r},t)=0$, and the
similar equation for the total local linear momentum density ${\bf
p}({\bf r},t)$, which includes the momentum flux density tensor, do
not involve the collision integral. Thus the shapes of the mean field
and of the nucleon effective mass, determined mostly by $n({\bf
r},t)$, are not directly affected by the two-body collisions.  However, the rate of 
the thermalization of the momentum distribution is controlled by the
two-body dissipation mechanism.  

 The NEDF should satisfy the local Galilean covariance, which implies that
 the total energy of the system, which is conserved,  
 can be represented as a sum~\cite{Engel:1975,Bender:2003,Bulgac:2013a}
\bea
&& E_\text{tot} \!\!=
E_\text{coll}(t)+E_\text{int}(t)
\equiv \int \!\!d{\bf r}\frac{ mn({\bf r},t){\bf v}^2({\bf r},t)}{2} \nonumber \\
&& +\int \!\!d{\bf r}\, {\cal E}\left (\tau({\bf r},t)-n({\bf r},t)m^2 {\bf v}^2({\bf r},t), n({\bf r},t),...\right ),\label{eq:etot}
 \eea
 where $n({\bf r},t)$ is the number density, $\tau({\bf r},t)$ is the kinetic density, and 
 ${\bf p}({\bf r},t)=mn({\bf r},t){\bf v}({\bf r},t)$ are linear 
 momentum and local collective/hydrodynamic velocity
 densities, and ellipses stand for various other densities.  
$\tfrac{{\bf p}({\bf r},t)}{n({\bf r},t)}$ is the position of the center of the local Fermi sphere in momentum space. 
 The first term in Eq.~\eqref{eq:etot} is the collective/hydrodynamic energy
 flow $E_\text{coll}$ and the second term is the intrinsic energy
 $E_\text{int}$ in the local rest frame.  For the sake of simplicity
 we have suppressed the spin and isospin DoF, even though they are included
 in all the actual calculations.
 The collective energy $E_\text{coll}(t)$ is not vanishing only in 
 the presence of currents and vanishes exactly for stationary states. 
 The inertia tensor in $E_\text{coll}(t)$ in the case of irrotational collective motion 
 is fully equivalent to the Werner-Wheeler inertial tensor~\cite{Pomorski:2012}.
 The intrinsic energy $E_\text{int}(t)$
 is determined only by the fermionic matter distribution.
 A similar partition of the total energy of the nucleus exists in the TDGCM 
 approach, see \cref{sec:II}.

\begin{figure}
\includegraphics[width=\columnwidth]{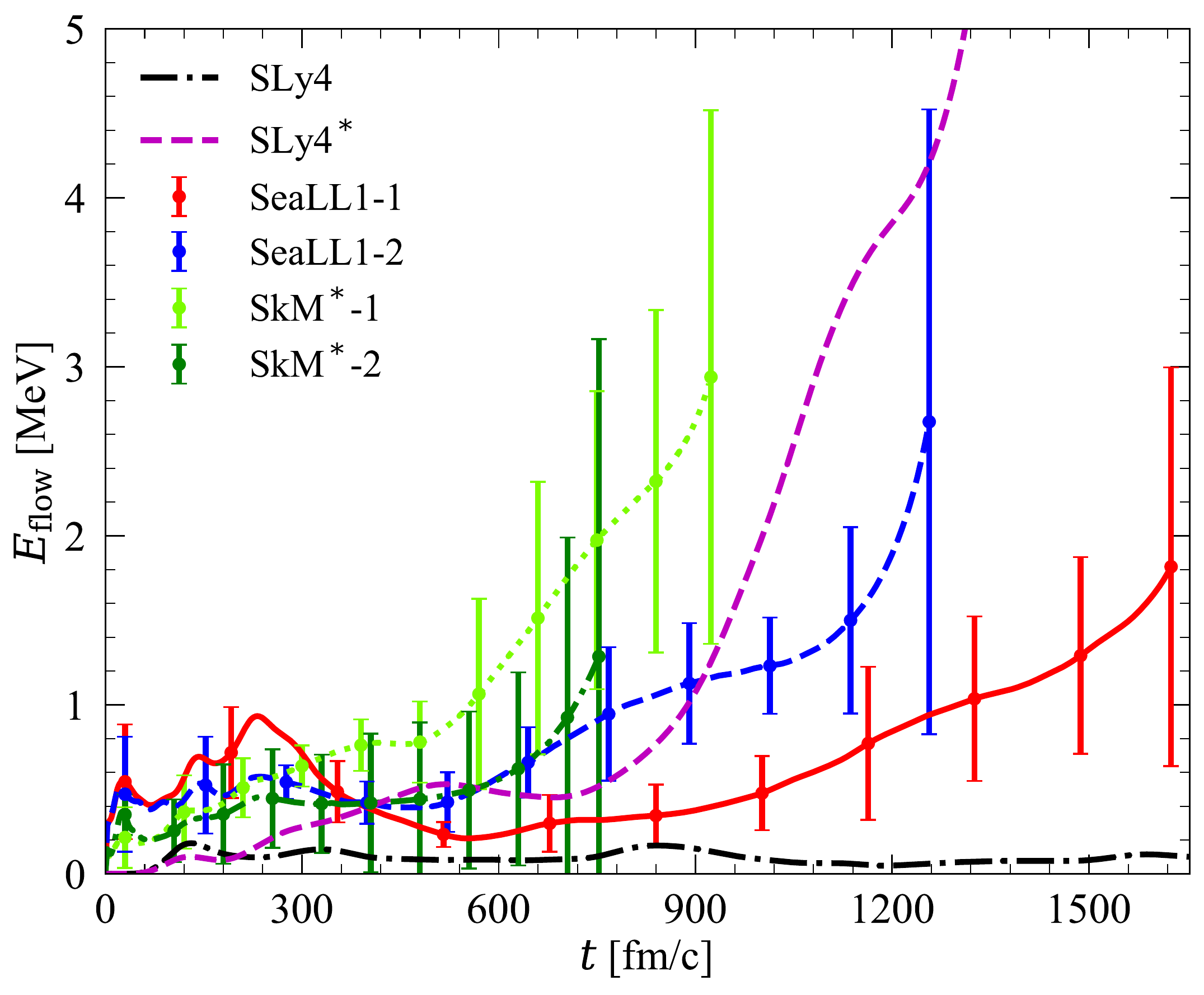}
\caption{
\label{fig:ecoll} (Color online)
The collective flow energy evaluated for NEDFs with
realistic pairing SLy4~\cite{Bulgac:2016} (dash-dot line), enhanced pairing SLy4* (dash line), and
for SkM*(dotted and dash-dot lines with error bars), and SeaLL1 
(solid and dashed lines with errors bars) sets.  The error bars illustrate the
size of the variations due to different initial conditions in case of SeaLL1-1,2 and SkM*-1,2 
NEDFs.  In the case of enhanced pairing 
NEDF Sly4* the time has been scaled by a factor of 1/10. }
\end{figure}

We have evaluated the collective flow
energy during the saddle-to-scission evolution, see
\cref{fig:ecoll},
\begin{align}\label{eq:ecoll}
E_\text{coll}(t) =  \int \!\!d {\bf r}\frac{ mn({\bf r},t){\bf v}^2({\bf r},t)}{2} 
= \int d{\bf r} \frac{ |{\bf p}({\bf r},t)|^2}{2m n({\bf r},t)},
\end{align}
which is a quantity unaffected by two-body collisions.
In the case of pure adiabatic evolution - as in TDGCM or ATDHF, see \cref{sec:II} -   
one expects a full conversion
of the collective potential energy into a collective flow energy of
$\approx 15\ldots 20$ MeV. 

\begin{figure}
\includegraphics[width=\columnwidth]{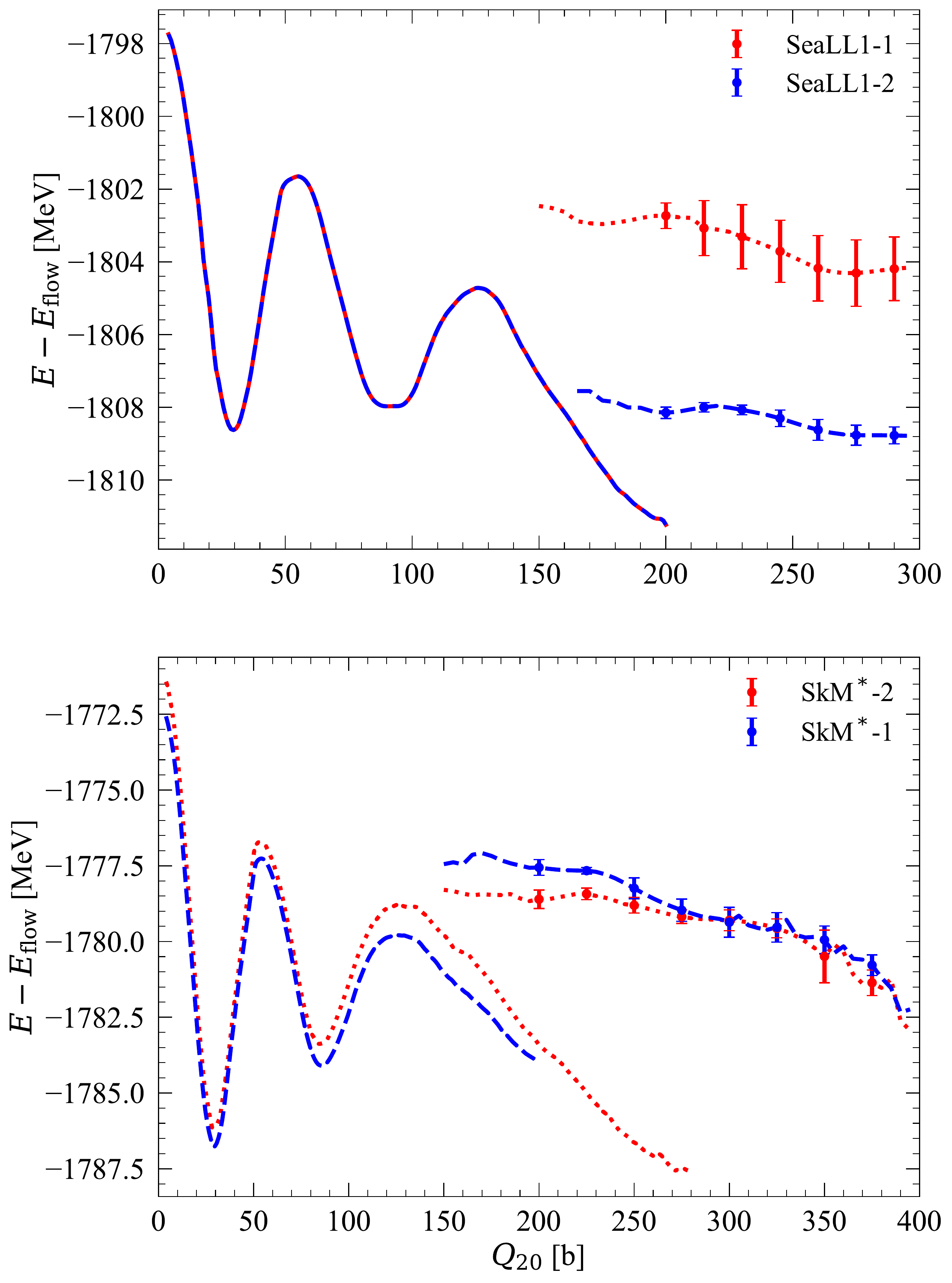}
\caption{
\label{fig:epot} (Color online)
The intrinsic energy $E_\text{int}(t)$ along the fission path 
for SeaLL1-1 (dotted line with error bars) and SeaLL1-2 in the upper panel, and 
SkM*-2 (dash line with error bars) and 
SkM*-1 (dotted line with error bars)  in the lower panel.
 The error bars illustrate the
size of the variations due to different initial conditions in case of SeaLL1-1,2 and SkM*-1,2 
NEDFs.
The collective potential energy determined in a constrained calculation, see Eq. \eqref{eq:U}
is represented in both panels with either dotted or dashed lines for the corresponding SeaLL1 and SkM* NEDFs. In the case 
of an adiabatic evolution along the fission path $E_\text{int}(t)$ would trace rather closely the 
collective potential energy determined in the constrained calculation. Scission configuration corresponds 
to a quadrupole momentum of the entire nuclear system $Q_{20}\approx 400$ b.}
\end{figure}

Surprisingly, our simulations point to an
unexpectedly small $E_\text{coll}$ from saddle-to-scission,
corresponding to a collective speed $\tfrac{v_\text{coll}}{c}\approx
0.002\cdots 0.004$, significantly smaller than the Fermi velocity
$\tfrac{v_F}{c}\approx 0.25$, see Figs. \ref{fig:ecoll} and \ref{fig:epot}.  Since in TDDFT one simulates the one-body
dynamics exactly, it is natural to discuss adiabaticity at the mean-field
level.  The transition rate between sp states is suppressed if the
time to cross an avoided level-crossing configuration satisfies the
restriction $ \Delta t\ll \tfrac{\hbar}{\Delta \epsilon} \approx 400$ fm/c,
where $\Delta \epsilon = \tfrac{1}{\rho_{\rm sp}(\epsilon_F)}$ is the average sp
energy level spacing at the Fermi level. Since on the way from
saddle-to-scission several dozen of avoided level crossings
occur~\cite{Barranco:1990,Bertsch:2018}, this condition is clearly
violated.  The collective motion is thus expected to be strongly
overdamped. From saddle-to-scission the nucleus behaves as a very
viscous fluid, the role of collective inertia is strongly suppressed,
and the trajectories follow predominantly the direction of the
steepest descent with the terminal velocity determined by the balance
between the friction and the driving conservative forces, see
\cref{fig:pes}.  

This result serves as the first microscopic
justification for the assumption  of the overdamped Brownian motion
model~\cite{Randrup:2011} and partially to the scission-point
model~\cite{Wilkins:1972,Wilkins:1976,Lemaitre:2015}.  
In both these phenomenological models it is assumed that the preformed FF 
are in thermal equilibrium and that the collective energy flow is either vanishing or very small. 
The main difference is that in the scission-point model there is no mechanism to ensure that 
all equilibrium scission configurations could be reached dynamically, 
while the nucleus evolves from the saddle-to-scission.  
It is equally surprising
that in the case of enhanced pairing, when the pairing condensates
retain their long-range order throughout the entire saddle-to-scission
evolution, the collective dynamics has the same general
characteristics.

The present results put a lower limit on the role of
the viscosity on fission times, as fluctuations can only lead to
longer trajectories~\cite{Bulgac:2019a}.  The character of the collective
dynamics unveiled here suggests that in physically realistic 
Langevin~\cite{Randrup:2011,Sierk:2017,Ishizuka:2017,Sadhukhan:2016,Sadhukhan:2017}
and TDGCM~\cite{Goutte:2005,Regnier:2016} studies the dynamics of the intrinsic DoF should  
be generated at (an approximately) fixed intrinsic energy,
since $E_\text{coll}(t)$ is small up to scission.   If the thermalization of the intrinsic DoF
is achieved fast enough and if the temperature of the system were constant the force driving the collective 
dynamics is determined by the free energy gradient~\cite{Frobrich:1998}
\beq
 {\bm F_{\bm Q}} =  -{\bm \nabla_{\bm Q}} [E_\text{int}({\bm Q},T)-TS({\bm Q},T)]\approx
{\bm \nabla_{\bm Q}} [TS({\bm Q},T)],
 \eeq
where $S({\bm Q},T)$ is the entropy and 
\beq
E_\text{tot} =
E_\text{int}({\bm Q},T)+E_\text{coll}(t) \approx E_\text{int}(t). \label{eq:eint}
\eeq  

However, as our results show the ``temperature'' of the nucleus, while 
descending from the saddle-to-scission increases, as $E_\text{int}(t)\approx \textrm{const.}$
In that case, for each set  of the collective variables 
${\bm Q}$ the temperature $T$ shall be adjusted so that
$E_\text{int}({\bm Q},T)$ remains practically equal to its starting
value, due to the smallness of $E_\text{coll}(t)$.  The intrinsic DoF
carry most of the intrinsic entropy of the fissioning nucleus and that drives
the fission dynamics until scission.  The entropy $S({\bf Q},T)$ and the temperature
are peaked along the bottom of the fission valley and there the free energy
${\cal F}({\bf Q},T)=E_\text{int}({\bm Q},T)-TS({\bm Q},T)$ reaches a minimum 
for fixed ${\bf Q}$ one expects to find the most 
probable fission path between the outer saddle and the scission configuration. 

In  order to include fluctuations one can proceed in at least two different ways. 
One possible avenue 
is to follow the procedure described in Ref.~\cite{Bulgac:2019a}. An alternative approach
is to introduce an appropriate number of collective variables ${\bm Q}$, minimize the grand 
canonical ensemble with respect to the sp DoF
\beq
\Omega = \min [E_\text{int}({\bm Q},T)-TS({\bm Q},T)-\mu_NN-\mu_ZZ-{\bm \lambda}\cdot{\bm Q} ]
\eeq	
and additionally vary the temperature until the constraint
\beq
E_\text{tot} \approx E_\text{int}({\bm Q},T)
\eeq
is satisfied, where $E_\text{tot}$ is the initial total energy of the fissioning nucleus.
At this point one would choose to make a step in the collective variable space
\beq
{\bm Q}\rightarrow {\bm Q}_\text{new}= {\bm Q}+\delta{\bm Q}
\eeq
and determining the new temperature $T_\text{new}$ as well,
by accepting or rejecting the new values ${\bm Q}_\text{new}$ according to the 
Metropolis criterion for the ratio between the level densities of the old $\rho({\bm Q})$ 
and new $\rho({\bm Q}_\text{new})$ configurations,  with probability 
\begin{align} 
P(Q\rightarrow Q_\text{new})&= \min\left [ \frac{\rho({\bm Q}_\text{new})}{\rho({\bm Q})},1\right ] \\
&\approx \min \left \{ \exp\left [ S({\bm Q}_\text{new},T_\text{new}) -S({\bm Q},T)  \right ], 1\right \}.  \nonumber
\end{align}

\begin{figure}[!htbp]
\includegraphics[width=\columnwidth]{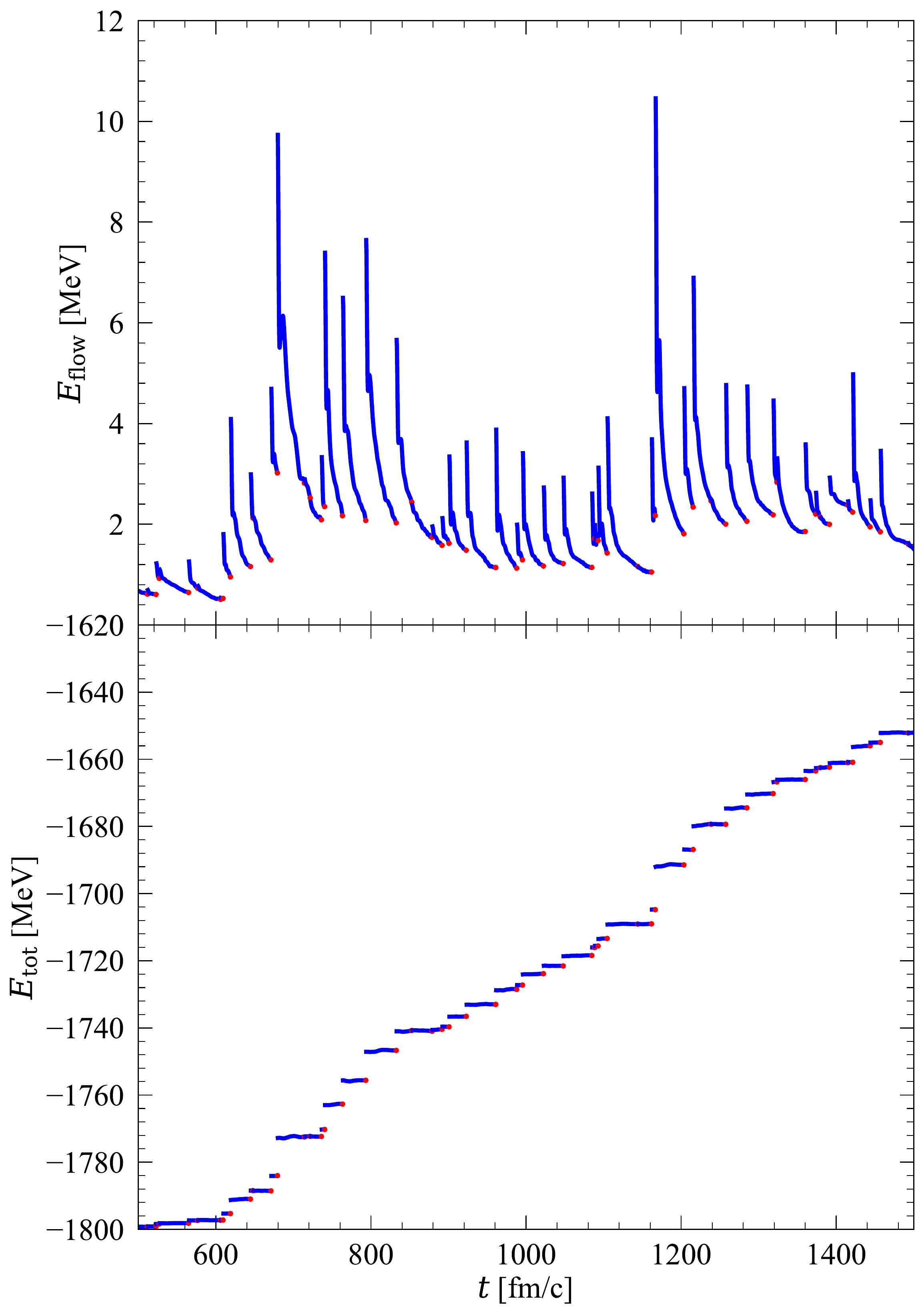}
\caption{
\label{fig:eflow_kicks} (Color online)
Upper panel: At times indicated with red dots we have applied collective quadrupole 
momentum kicks to both neutrons and protons, see \cref{eq:kicks}, with random 
values of $\eta$.
Lower panel: The time evolution of the total energy of the nucleus, 
in the rest frame of the nucleus, after we have applied 
collective kicks to both neutrons and protons with random 
values of $\eta$.
}
\end{figure}

\subsection{Relaxation of the collective degrees of freedom} \label{sec:V C}

To demonstrate the overdamped character of the fission dynamics, we 
performed the following theoretical experiment. We have applied 
at random times, the red dots in \cref{fig:eflow_kicks},
collective kicks to the nucleus of random intensities $\eta$ according to the prescription
\begin{align}
\label{eq:kicks}
\begin{pmatrix}                                          
u_{k\sigma} (\vect{r},t) \\
v_{k\sigma} (\vect{r},t)
\end{pmatrix}
\rightarrow 
\begin{pmatrix}
\exp [ +i\eta\phi({\bf r})]   u_{k\sigma} (\vect{r},t) \\
\exp [ -i\eta\phi({\bf r})]   v_{k\sigma} (\vect{r},t)
\end{pmatrix},
\end{align} 
where $\sigma = \uparrow,\downarrow$,
which immediately resulted in an increase of the collective flow energy only. The momenta of all nucleons
are instantaneously shifted by the same amount $\Delta {\bf p} = \hbar \eta {\bm \nabla}\phi({\bf r})$ and 
the excitation energy injected into the nucleus by a such a collective kick is 
\beq
\Delta E = \int \!\!\!\!d{\bf r} n({\bf r})\left [
\frac{ |\eta \hbar {\bm \nabla}\phi({\bf r})+m{\bf v}({\bf r},t)\ |^2}{2m} - \frac{ m|{\bf v}({\bf r},t)\ |^2}{2}\right ],
\eeq 
where $n({\bf r})$ is the total number density and ${\bf v}({\bf r},t)$ 
is the local collective velocity prior to the kick. Immediately 
after such a kick the density distribution has the same profile as just before the kick, 
as the phase of the qpwfs do not affect number densities, but affect the currents and the kinetic energy density.
After a relatively short 
time, of the order of a few 10's fm/c, this excess collective flow energy is rapidly dissipated
into intrinsic DoF and the nucleus is thus heated up. This added energy is never returned into 
the collective flow energy of the fissioning nucleus. 
After each collective ``kick'' the intrinsic energy of the nucleus increases, see \cref{fig:eflow_kicks}. 
Even though  the intrinsic energy increased by $\approx$ 150 MeV after many collective ``kicks'',
the rate at which the additional energy in absorbed does not visibly change.
This serves as an additional argument that the 
one-body dissipation mechanism is very effective in bringing the collective flow 
velocity to the terminal velocity, which is achieved when the friction force 
cancels the driving force, see also the discussion in  \cref{app:C}.

\begin{figure}[!htbp]
\includegraphics[width=\columnwidth]{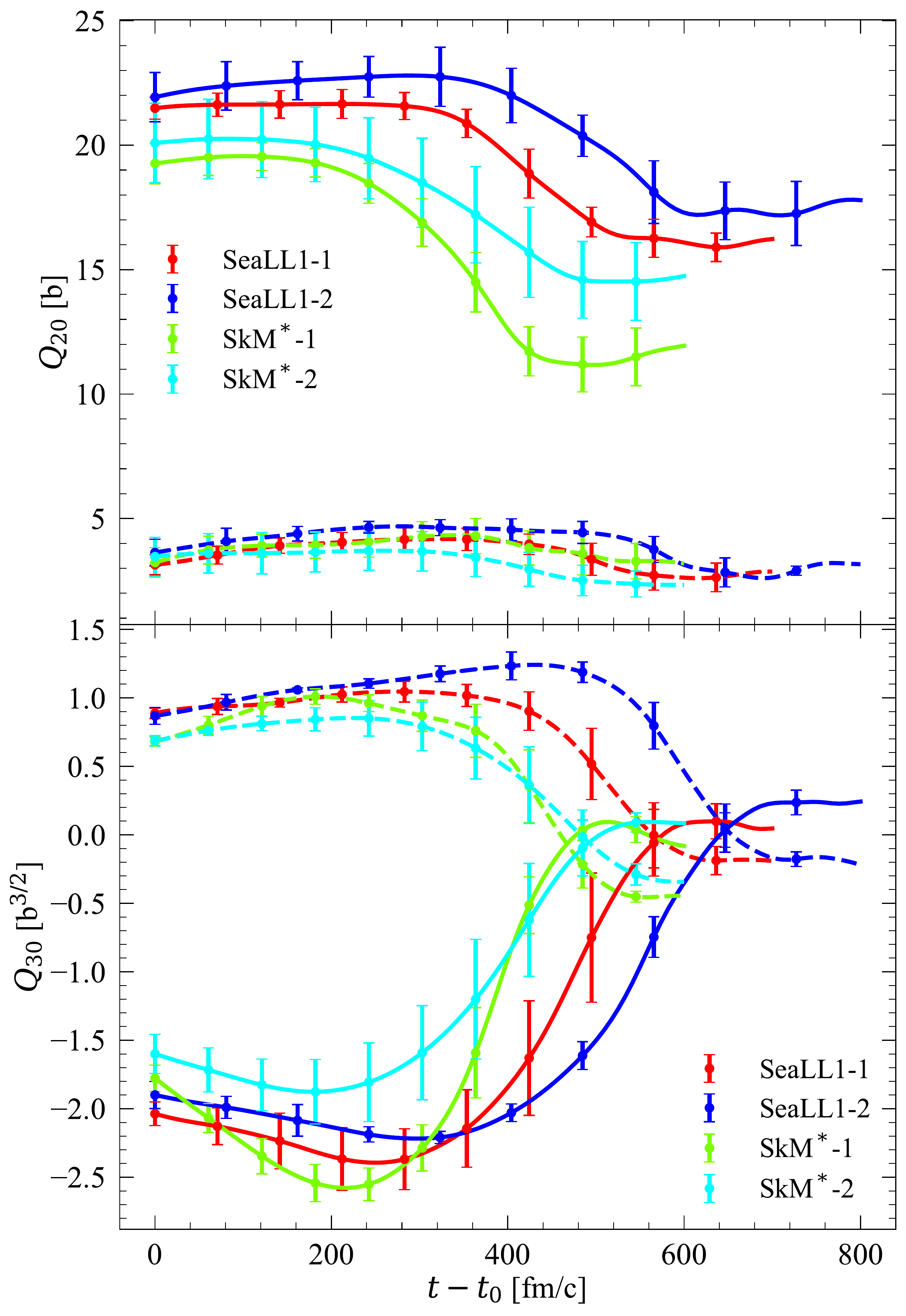}
\caption{\label{fig:q2-q3} (Color online)
The evolution of the quadrupole $Q_{20}$ and octupole $Q_{30}$ 
moments of the light (solid lines) and heavy (dashed lines) FFs 
before and after scission. The time $t_0$ stands for 
the moment when the distance between the two fragments is about 15 fm and 
the neck of mother nucleus is formed. Scission occurs at $t-t_0 \approx 300$ fm/c.
The solid lines represent the multipoles moments of the light fragment 
and the dashed lines the multipole moments of the heavy 
fragment in the case of the SeaLL1-1,2 and SkM*-1-2 respectively.
The error bars illustrate the
size of the variations due to different initial conditions in case of SeaLL1-1,2 and SkM*-1,2 
NEDFs.
 }
\end{figure}

\begin{figure}[!htbp]
\includegraphics[width=\columnwidth]{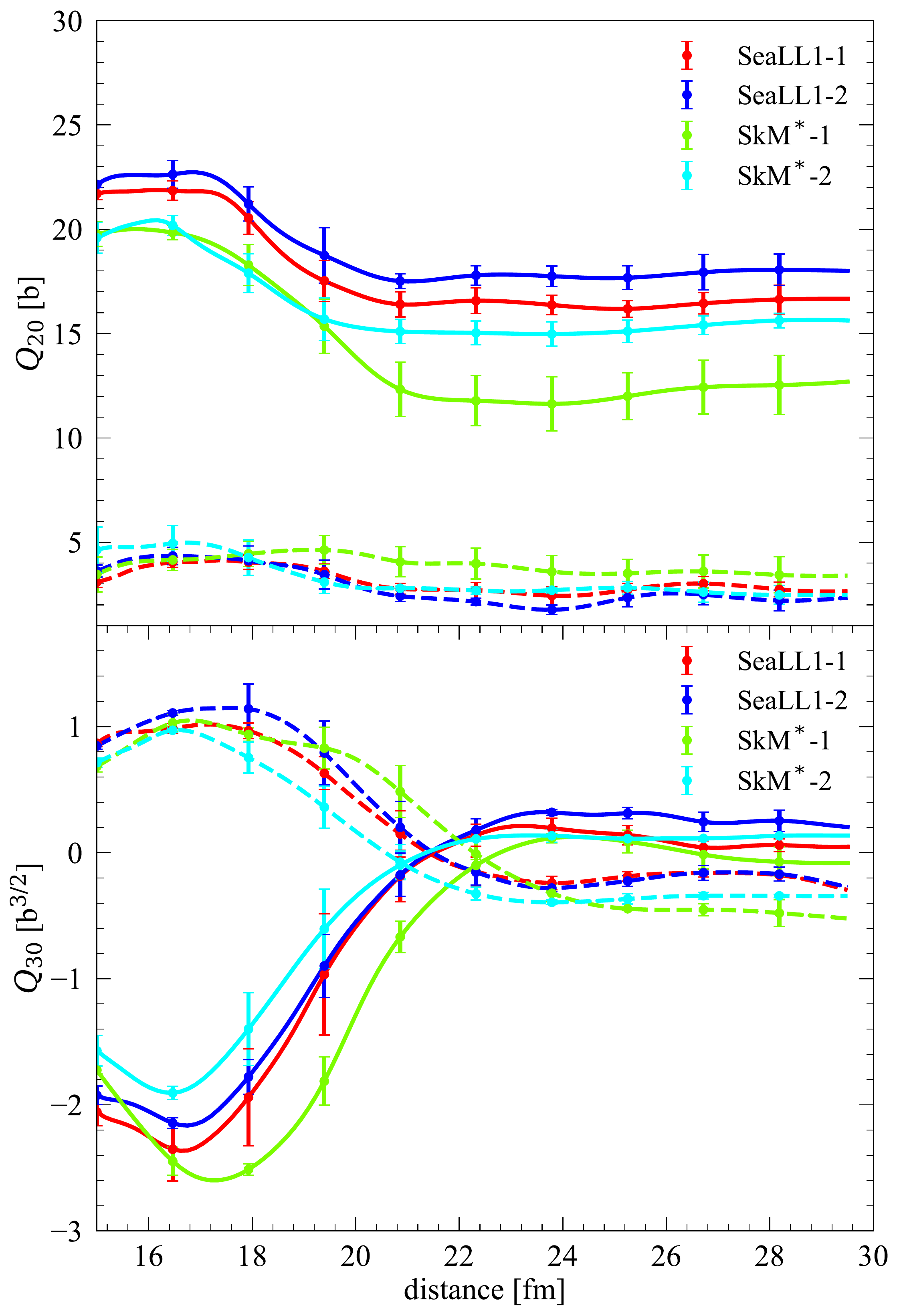}
\caption{\label{fig:q2-q3-d} (Color online)
The evolution of the quadrupole $Q_{20}$ and octupole $Q_{30}$ 
moments of the light (solid lines) and heavy (dashed lines) FFs 
before and after scission, here as a function of the separation between the FFs. At scission
the separation between the FFs is $\approx$ 17 fm. The meaning of various 
lines and error bars are explained in the caption to Fig. \ref{fig:q2-q3}.}
\end{figure}

\subsection{Shape relaxation of fission fragments }  \label{sec:V D}

The one-body dissipation is important both before and after scission. The light fragment at scission is very elongated 
and both fragments have also a noticeable amount of octupole deformation, very different than the 
corresponding moments in the ground state. In \cref{fig:q2-q3} we show the evolution of 
$Q_{20}$ and $Q_{30}$ for both FFs after scission. All these moments relaxed rather rapidly, without 
performing any oscillations to the values very close to the ground state values. 
Remember however that both FFs are not cold. The absence of shape 
oscillations is another strong indication that one-body dissipation is strong and  
that even the individual FFs LACM is overdamped always. The relatively 
large quadrupole deformation energy of the light fragment is thus converted into heat and 
its quadrupole moment is considerably reduced. Both fragments are octupole deformed 
at scission, and these octupole moments relax to relatively small values. As the deformation 
energy is converted into intrinsic excitation energy the temperatures of both FFs increase, 
compared to the corresponding temperatures at scission.
In \cref{fig:q2-q3-d} we plot the evolution of these quadrupole and octupole moments as a function of the separation 
between the FFs. The FFs achieve their relaxed shapes  at a separation between FFs $\approx 22$ fm, 
thus when the distance between their tips is about 4 to 5 fm. This FFs separation at which their shapes are relaxed 
is noticeably larger than the separation 
considered in the scission point model~\cite{Wilkins:1972,Wilkins:1976,Lemaitre:2015} or in the Brownian 
motion model~\cite{Randrup:2011,Randrup:2011a,Randrup:2013,Ward:2017,Randrup:2018}.

\begin{figure}[!htbp]
\includegraphics[width=\columnwidth]{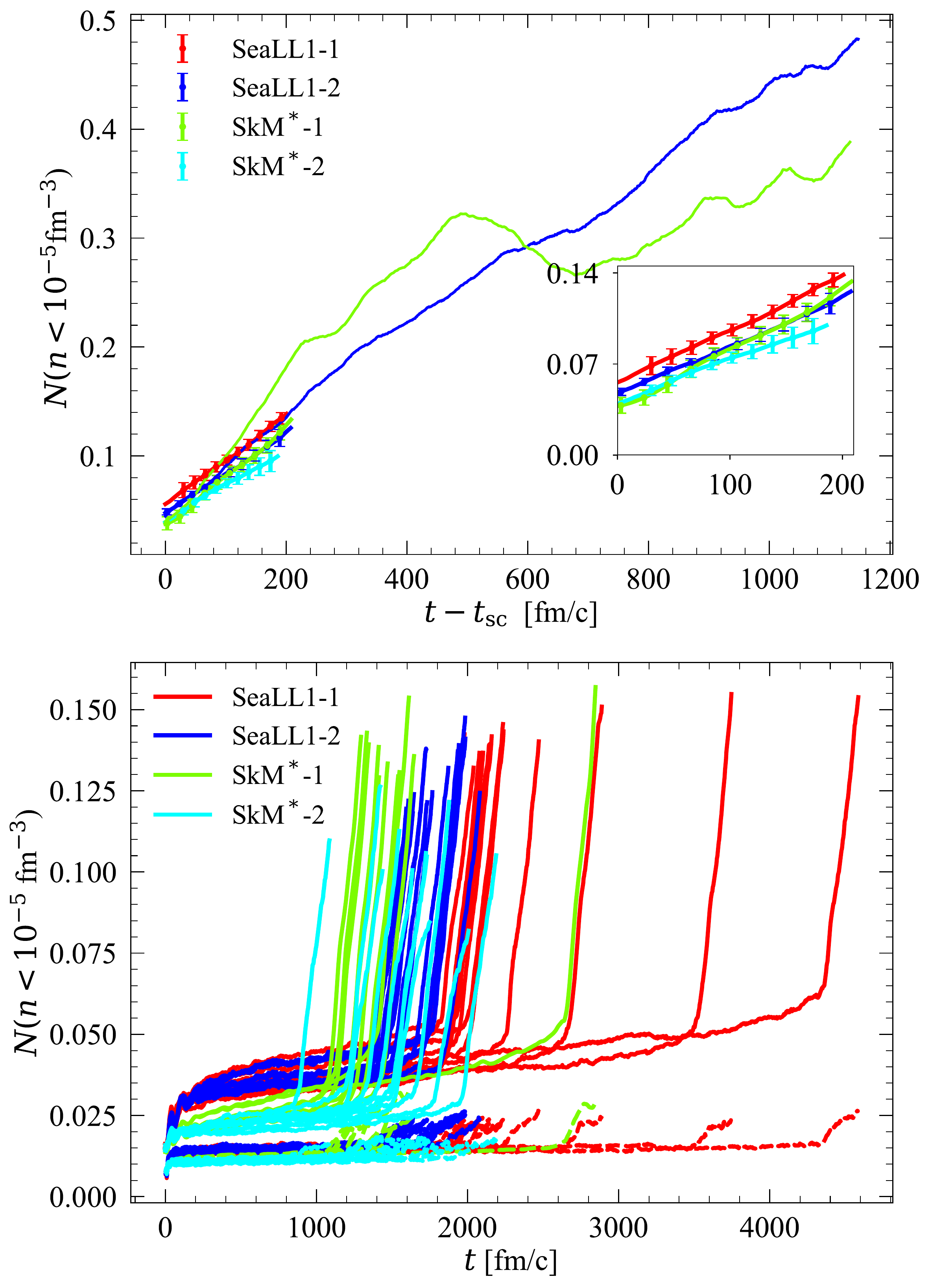}
\caption{\label{fig:np_scission} (Color online)
Upper panel: The number of neutrons emitted predominantly after scission. 
The error bars quantify the size of the fluctuations between 
trajectories corresponding to different initial conditions; Lower panel:
the number of neutrons as a function of time emitted in each trajectories in the inset of upper one.
}
\end{figure}

\subsection{Neutrons emitted by fission fragments} \label{sec V C}

Neutron emission is an important mechanism in the de-excitation of FF.
We have also estimated the neutron emission rates before the FFs are fully
separated, $ 4.0\times 10^{-4}$ neutrons/(fm/c),
which are rather stable with respect to the variation of initial
conditions, deformation, initial energy, or NEDF.  By the time the FFs
reach a separation of $\approx 60$ fm we find that about 0.4 neutrons
are emitted on average, in fair agreement with theoretical
estimates~\cite{Capote:2016a,Carjan:2019} 
and with experimental hints of neutrons emitted before full 
acceleration~\cite{Vorobyev:2010,Kornilov:2001,Kornilov:2001a,Petrov:2005}.

In \cref{fig:np_scission} 
we show show the number of neutrons in the volume 
where the number density $n(\vect{r},t) \le 10^{-5}$ fm$^{-3}$ as a function of time after 
scission for all the fission trajectories we have evaluated. The neutron emission rates 
(the slopes) demonstrate a robust independence on the initial conditions or the NEDF used.
These neutrons are emitted
preferentially parallel to the FFs motion, a conclusion likely
affected by the finite transverse size of simulation boxes.

\section{Conclusions}

The present study is the first in literature in which no restrictions have been 
imposed on the mean field dynamics and the pairing field was treated beyond the BCS approximation. 
In TDHF+BCS treatments reported so far in literature a relatively low-energy 
cutoff was always imposed on the energy band where pairing correlations are active. 
Arguments are often presented that such a limited space for pairing correlation is 
sufficient, as the energy gain is reproduced with enough accuracy.  However, as Anderson 
notes~\cite{Anderson:1984} (see page 128), when he discusses the ``Quantum Chemists' Fallacy No. 1 and 2,''
of which even Wigner was partially guilty, ``you may get 
pretty good energetics out of a qualitatively wrong state.''
The perfect example is the case of a superconductor, in which in spite of the fact that the 
contribution from the condensation energy is negligible, the 
wave function with pairing correlations leads to qualitative changes. 

It is  well known that pairing correlations in nuclei, unlike in 
superconductors, are mostly due to short range attractive forces, and that is 
inconsistent with the assumption that pairing correlations 
are active in a narrow energy window only.
In particular this is also true for the popular finite range Gogny interaction~\cite{Decharge:1980}, 
in which case the pairing cutoff energy could be as high as ${\cal O}(100)$ MeV.   
The contribution of the pairing correlations to the total energy of the nucleus is very modest 
and by varying the energy window
where these correlations are active is not going to influence noticeably the total energy energy. However,  the 
total nuclear wave function is affected significantly by the size of the 
energy window in which the pairing correlations are allowed to participate, particularly in dynamics.
As we do not implement any arbitrary energy cutoff on pairing correlations, 
we are able to perform an accurate microscopic test on 
whether the LACM fission dynamics is indeed adiabatic in character. 

Adiabaticity in LACM is typically conflated with slowness of 
the collective motion~\cite{Baranger:1972,Baranger:1978,Villars:1978,Ring:2004,Pomorski:2012,Schunck:2016}, 
an assumption which allows one to  introduce collective DoF which are decoupled from the intrinsics DoF
and also legitimizes the introduction of a collective Hamiltonian.
Until now the validity of this assumption has not been checked, since the 
required simulations exceeded the computer capabilities of previous researchers. 
The only previous serious attempt we are aware of is due to \textcite{Ledergerbert:1977}, 
who however were unable to arrive at a conclusive decision, basically because the 
phase space they were able to consider at the time, was incontrovertibly  too small.

The one-body dissipation mechanism 
has been suggested by \textcite{Blocki:1978} and almost three decades earlier 
by Fermi~\cite{Fermi:1949}, under a different name, 
in order to explain the energy  spectrum of cosmic rays. Fermi suggested that charged particles (protons)  collide 
with moving magnetic fields  and as a result are accelerated to very large energies. 
The energy from the moving magnetic fields, which play a similar role to the moving nuclear 
surface, is transferred to the charged particles (protons), which play a similar role to the intrinsic 
nucleonic DoF.   In a nucleus, which  is a finite system, the intrinsic DoF bring the nuclear walls
to an almost standstill as present results manifestly demonstrate. 

One-body dissipation is automatically incorporated in TDDFT and it was 
conjectured for many years to be important in LACM. \textcite{Negele:1978} 
concluded that one-body dissipation is important in fission dynamics, based 
however on studying only three TDHF trajectories for fission of $^{238}$U, 
with three values for a \emph{static} (sic) pairing field with the  pairing  gap $\Delta= 0.7$ , 2.0, and 6.0 MeV
and comparing these results with classical Langevin simulation of fission dynamics, 
with one-body dissipation estimated for a Fermi gas model. 
Notably, the kinetic energy of the FFs at scission determined in this TDHF study was about 11-12 MeV, 
an order of magnitude larger than the value we determine, and also a value consistent with adiabatic 
character of the large amplitude collective motion from saddle-to-scission, which we dispute here.
From this type of comparison 
these authors concluded that one-body dissipation  is important in fission dynamics. 
Over the years the practitioners of the Langevin type of simulations have claimed 
that both one-body and two-body dissipation are important in low energy fission 
dynamics~\cite{Frobrich:1998,Sierk:2017,Ishizuka:2017,Sadhukhan:2016,Sadhukhan:2017}, 
using phenomenologically adjusted dissipation coefficients, though it is not 
always clear how one can disentangle the two forms of dissipation, 
see e.g. the study performed by \textcite{Sierk:2017}.
Not all Langevin implementations are compatible with one another, 
though the level of agreement with data is about the same. Moreover there 
are TDGCM simulations~\cite{Regnier:2016,Regnier:2019}, in which the nature of fluctuations is qualitatively 
different, quantum in character,  but the quality of the agreement with experiment
for FFs mass yields are comparable with those achieved in Langevin approaches, 
where the fluctuations are thermal in character. 
All this might suggest that the FFs mass yields are likely
not very sensitive to the nature and details of various fluctuations models. 

Only in the last few years Randrup and collaborators~\cite{Randrup:2011,Randrup:2011a,
Randrup:2013,Ward:2017,Randrup:2018} took this assumption to the extreme and 
suggested that fission dynamics is actually overdamped and one should replace 
the Langevin approach with the Smoluchowski approach. In this case the role of the 
collective inertia becomes irrelevant and a more accurate description
of the fission yields, the TKE, and of the sharing of the TXE can be achieved.  This
approach shares similarities with the scission point 
model~\cite{Wilkins:1972,Wilkins:1976,Lemaitre:2015}, as both
approaches assume total conversion of the saddle-to-scission collective potential energy difference 
into the FFs intrinsic excitation energy or, in other words, into heat. 
As we demonstrate here
however, these phenomenological models neglect the fact FFs fragments attain 
their equilibrium shapes when they are quite well separated and thus they underestimate the 
excitation energy of the FFs. From the scission configuration to the point where 
the FFs reach their equilibrium shapes the deformation energy of the FFs is converted into 
additional  internal excitation energy.

We have shown in \cref{sec:V C} that at scission both FFs emerge with significant 
quadrupole and octupole deformations, which relax to their equilibrium values only after the FFs are 
significantly separated, see Figs. \ref{fig:q2-q3} and \ref{fig:q2-q3-d} below.  
The ``surplus'' deformation energies of the FFs  are converted
into additional internal excitation energies of the FFs, on top of their ``thermal'' excitation energy
achieved during the descend from saddle-to-scission, and it affects the average neutron
multiplicity spectra, which are emitted by the fully accelerated FFs, see Fig. \ref{fig:nubar}. 
No phenomenological models so far incorporated these aspects, particularly the octupole deformations 
of the emerging FFs at scission. As a result these phenomenological models
underestimate the magnitude the excitation energies of the FFs and consider only an
oversimplified excitation energy mechanism between FFs.

If dissipation is important and moreover overdamped in fission dynamics, the introduction of collective DoF freedom, 
of a potential energy surface and of a collective inertia, and the decoupling of the DoF 
into collective and intrinsic become highly questionable. We have establish that
one-body dissipation is very strong and that in fission LACM 
is strongly overdamped under almost any reasonable assumptions. Under such circumstances 
the role of ``collective inertia'' becomes irrelevant and the theoretical arguments in favor of a
TDGCM/ATDHF approach or Langevin approach become questionable as well, along with
the mere definition of collective variables, as a collective Hamiltonian by definition 
describes a non-dissipative motion.

We have determined that the memory of the initial conditions near the outer barrier are 
rather quickly forgotten, in a relaxation time $\tau_\text{relax}\approx {\cal O}(50)$ fm/c. As the saddle-to-scission is 
$\tau_{s\rightarrow s}\approx {\cal O}(10^3 \ldots 10^4)$ fm/c, the widths of the 
FFs mass, charge, TKE, TXE, and spin distributions  are determined during this relatively fast
non-equilibrium evolution interval from saddle-to-scission, due 
to the presence of non-negligible fluctuations. During this interval of time 
practically the entire gain in potential energy of a nucleus sliding down from the saddle-to-scission
is converted into internal heat. 

We have used three different NEDFs (SLy4, SeaLL1, and SkM*), along with variations in 
treatment of  the pairing correlations, and our conclusions are quite robust.  All these NEDFs
satisfy basic constraints: the nuclear matter is liquid in nature (thus mostly 
incompressible) and characterized by a significant surface tension and
isospin asymmetry due to a strong Coulomb interaction, and the spin-orbit and the pairing correlations have realistic values. When 
all these basic requirements are satisfied the emerging most likely values for the 
TKE, atomic, and charge numbers are in agreement with experiment without 
the need of any additional fitting. In addition we were able to extract the excitation 
energy sharing between the FFs. Depending on the initial energy of the fissioning nucleus 
the excitation energy sharing changes in a manner which appears to be in agreement 
with experimental data on average neutron multiplicities.

Many other quantities of interest 
for various applications could also be extracted, such as the angular momentum 
distributions of the FFs and their parities~\cite{Bertsch:2019,Bulgac:2019x}.

The only disadvantage of the present approach is that it lacks 
fluctuations, which likely could be later added~\cite{ Bulgac:2019a}. Upon introductions of fluctuations one 
should be able extract FFs mass, charge, TKE, and TXE  distributions.

\begin{acknowledgments}
We are grateful to G. F.~Bertsch for numerous discussions and
helpful comments on the manuscript and to P. Magierski for a number
of suggestions to improve the narrative. We thank I. Abdurrahman for helping  preparing \cref{fig:Eq}.

The work of AB and SJ was supported by U.S. Department of Energy,
Office of Science, Grant No. DE-FG02-97ER41014 and in part by NNSA
cooperative agreement DE-NA0003841.  The TDDFT calculations have been
performed by SJ at the OLCF Summit, Titan, and Piz Daint and for generating
initial configurations for direct input into the TDDFT code at OLCF
Titan and Summit and NERSC Edison. This research used resources of the Oak Ridge
Leadership Computing Facility, which is a U.S. DOE Office of Science
User Facility supported under Contract No. DE- AC05-00OR22725 and of
the National Energy Research Scientific computing Center, which is
supported by the Office of Science of the U.S. Department of Energy
under Contract No. DE-AC02-05CH11231.  We acknowledge PRACE for
awarding us access to resource Piz Daint based at the Swiss National
Supercomputing Centre (CSCS), decision No. 2016153479.

The work of KJR is supported by US DOE Office of Advanced Scientific
Computing Research and was conducted at Pacific Northwest National
Laboratory and University of Washington.
  
The work of NS was supported by the Scientific Discovery through
Advanced Computing (SciDAC) program funded by the U.S.~Department of
Energy, Office of Science, Advanced Scientific Computing Research and
Nuclear Physics, and it was partly performed under the auspices of the
US Department of Energy by the Lawrence Livermore National Laboratory
under Contract DE-AC52-07NA27344.  NS performed the calculations of
the initial configurations and of the temperatures of the FFs.  Some of the calculations reported
here have been performed with computing support from the Lawrence
Livermore National Laboratory (LLNL) Institutional Computing Grand
Challenge program.

The work of IS was supported by the US Department of Energy through the 
Los Alamos National Laboratory. Los Alamos National Laboratory is operated 
by Triad National Security, LLC, for the National Nuclear Security Administration 
of U.S. Department of Energy (Contract No. 89233218CNA000001).

  \end{acknowledgments}


\appendix

\section{Numerical implementation} \label{app:A}

In the present study we have significantly increased the size of the simulation 
box compared to our proof-of-principle results of \cite{Bulgac:2016}, from
$22.5^2\times 40\; \text{fm}^3$ to $30^2\times 60\; \text{fm}^3$, using the 
same lattice constant $l= 1.25$ fm, which corresponds to a momentum cutoff 
$p_c= \tfrac{\hbar \pi}{l} \approx 500$ MeV/c in each spatial direction. The momentum 
space is in this case a cube with volume  $(2p_c)^3$. The total number of 
available quantum sp states is thus
\beq
{\cal N}_{qs} =4 \frac{N_xN_yN_z l^3 \times (2p_c)^2}{(2\pi \hbar)^3} = 4N_xN_yN_z=110\,592,
\eeq
where the factor 4 accounts for the spin and isospin DoF.
This significant increase in the size of the 
calculations was required in order to ensure numerical stability 
in larger spatial lattices~\cite{Grineviciute:2018}.
We use Fast Fourier Transform (FFT) to 
compute spatial derivatives, since it reduces the number of floating point 
operations significantly, while practically ensuring machine precision for 
derivatives. We avoid computing first-order derivatives when possible, e.g., we 
take advantages of standard relationships such as
\bea
&& \bm{\nabla} F({\bm r})\cdot \bm{\nabla} G({\bm r}) \rightarrow \\
&& \frac{1}{2}
\Big( 
\Delta [ F({\bm r}) G({\bm r}) ] - \Delta [F({\bm r})] G({\bm r})  -F({\bm r})\Delta [G({\bm r}) ]
\Big), \nonumber
\eea 
for increased numerical accuracy. The evaluation of first order derivatives requires the elimination 
of the highest frequency in the Fourier transform for numerical accuracy. If  couplings to gauge fields is 
required, as in Ref.~\cite{Stetcu:2014}, or when evaluating terms linear in momentum,
we use the discretized symmetrized form
\bea
&& \bm{A}({r},t)\cdot \bm{\nabla}\psi({\bm{r},t}) \rightarrow  \\ 
&& \frac{1}{2}
\left [ \bm{A}({r},t)\cdot \bm{\nabla}\psi({\bm{r},t}) 
     + \bm{\nabla}\psi({\bm{r},t}) \cdot   \bm{A}({r},t) \right  ]. \nonumber
\eea
When evaluating first order derivatives of products of functions we use Leibniz rule 
\beq
\bm{\nabla} \left [A(\bm{r})B(\bm{r})\right ] \rightarrow \
 B(\bm{r})  \bm{\nabla} A(\bm{r})+  A(\bm{r}) \bm{\nabla} B(\bm{r}).
 \eeq
The use of this rule is particularly important to ensure numerically accurate gauge invariance.
As discussed in Refs. \cite{Bulgac:2013,Ryssens:2015} with a 
careful choice of the size of the box and of the spatial lattice constant one can achieve very high 
numerical accuracy with relatively large values of lattice constant $l$. In computing the Coulomb 
potential we use the method described in Ref. \cite{Castro:2003} to solve the Poisson equation 
in order to eliminate the contributions 
from images, which  are inherent when using periodic boundary conditions.  

The initial state is generated using the code HFBTHO~\cite{Perez:2017} with 
appropriate constraints on the expectation value of the quadrupole  
$\langle\hat{Q}_{20}\rangle$ and octupole $\langle\hat{Q}_{30}\rangle$ 
moments. HFBTHO calculations are performed in a stretched basis of 
$N_{0} = 28$ shells with the deformation $\beta$ and the oscillator frequency 
$\omega_0$ set as in Ref.\cite{Schunck:2014}. 
The matrix of the Bogoliubov transformation is then transformed 
in the coordinate space representation on a spatial lattice of size 
$N_xN_yN_z \times N_xN_yN_z$ according to
\begin{align}
\begin{pmatrix}
u_{k\sigma}({\bm r}) \\
v_{k\sigma}({\bm r}) 
\end{pmatrix} =
\begin{pmatrix} 
\sum_{n} U_{nk}\psi_{n\sigma}({\bm r}) \\
\sum_{n} V_{nk}\psi_{n\sigma}^{*}({\bm r})
\end{pmatrix},
\end{align}
where $\psi_{n\sigma}({\bm r})$ are the harmonic oscillator basis spinors, see 
\cite{vautherin1973,stoitsov2005}. The qpwfs are used to reconstruct the 
densities and the potentials, with the help of which we construct the
initial conditions in Eq.~\eqref{eq:tdslda} (in  total $4N_xN_yN_z$ neutron and proton qpwfs)
by diagonalizing the full pairing Hamiltonian matrix, including the constraints
\begin{align}
\begin{pmatrix}
h_{\uparrow \uparrow} -q & h_{\uparrow \downarrow} & 0 & \Delta \\
h_{\downarrow \uparrow} & h_{\downarrow \downarrow} -q& -\Delta & 0 \\
0 & -\Delta^* &  -h^*_{\uparrow \uparrow}  +q& -h^*_{\uparrow \downarrow} \\
\Delta^* & 0 & -h^*_{\downarrow \uparrow} & -h^*_{\downarrow \downarrow} +q
\end{pmatrix}
\begin{pmatrix}
u_{k\uparrow} \\
u_{k\downarrow} \\
v_{k\uparrow} \\
v_{k\downarrow}
\end{pmatrix} = E_k
\begin{pmatrix}
u_{k\uparrow}  \\
u_{k\downarrow} \\
v_{k\uparrow} \\
v_{k\downarrow}
\end{pmatrix}, \label{eq:eqp_static}
\end{align}
where $q=\sum_l \lambda_lQ_{l0} $ stands for all constraints, including the corresponding Lagrange multipliers.

A larger simulation box allows us to more precisely characterize the FFs 
properties. The sp level density is denser, and during the 
dynamics sp states mix more easily. This forced us to include all
sp levels in the simulations, in order to avoid numerical 
instabilities for long-time trajectories, due to level crossings. This also led to a slightly modified 
renormalization procedure of the pairing gap constant, using Eqs. (5.47-5.50) 
from Ref.~\cite{Castin:2012} in monograph~\cite{Zwerger:2011}, in a similar manner to what was described in
Refs. \cite{Bulgac:2002,Yu:2003,Bulgac:2011a}
\beq
\frac{1}{g_\text{eff}}=\frac{1}{g}-\frac{2.442 \,m}{4\pi \hbar^2 l},
\eeq
where $l$ is the lattice constant.
The number of coupled nonlinear 
time-dependent 3D partial differential equations (PDEs) solved increased 
significantly from $\approx$ 56\,000 in Ref.~\cite{Bulgac:2016} to 
$16\times N_xN_yN_z =$ 442\,368 PDEs for a typical lattice $N_xN_yN_z=24^2\times 48$. While evaluating the neutron emission 
rates, we have used in a couple of instances an even larger simulation box 
$30^2\times120\; \text{fm}^3$, which amounted to evolving in time 884\,736 PDEs. 

The larger cut-off energy and the larger number of PDEs required a smaller 
time-step integration $\Delta t=0.03$ fm/c, leading to an error $\epsilon 
\sim (E_{\rm cut}\Delta t/\hbar)^6$, which is required per time-step 
for the predictor, modifier, and corrector steps, and where the maximum 
$E_{\rm cut}= (3p_c^2/2m)$. We use the Adams-Bashforth-Milne 
predictor-modifier-corrector time-integration algorithm~\cite{Hamming}:
\bea 
&&p_{n+1}=\frac{ y_{n}+y_{n-1} }{2}  \\
&&+\frac{h}{48} \left (119 {\bf y}'_{n} - 99y'_{n-1}  + 69y'_{n-2}-17y'_{n-3} \right ) +\frac{161}{480}h^5y^{(5)}, \nonumber \\
&& m_{n+1}=p_{n+1}-\frac{161}{170}\left ( p_{n}-c_{n}\right ) +\frac{923}{2880}h^6y^{(6)},\\
&&c_{n+1}= \frac{ y_{n}+y_{n-1} }{2}  \\
&&+\frac{h}{48}\left ( 17 {\bf m}'_{n+1}+51y'_{n}+3y'_{n-1}+y'_{n-2}\right )  - \frac{9}{480}h^5y^{(5)}, \nonumber\\
&& y_{n+1}=c_{n+1} +\frac{9}{170}\left (p_{n+1}-c_{n+1}\right ) -\frac{43}{2880}h^6y^{(6)},
\eea
where we have indicated with a bold symbol the only two times during a time-step when 
the right hand side of the differential equation has to be evaluated, namely ${\bf y}'_{n}$ for $p_{n+1}$ 
and ${\bf m}'_{n+1}$ for $c_{n+1}$.
The scheme
has an overall accuracy of $\epsilon \sim (E_{\rm cut}\Delta t /\hbar)^6=5\times 10^{-10}$ per 
time-step. It also requires only two evaluations of the right-hand-side 
of Eq.~\eqref{eq:tdslda}, namely only two applications of the Hamiltonian on 
the qpwfs. To start the propagation, and to check-point restart a previous 
calculation, we use a Taylor expansion of the unitary mean field propagator.
With such a time-step the particle number is conserved practically with 
machine precision and the total energy with an error 
$\approx 0.25\cdots 0.5$ MeV  for trajectories as long as 3,000 fm/c, 
thus 100,000 time-steps, see Table \ref{table:accuracy}. Our codes use double precision. 
In starting calculations or after checkpoint/restart we use for the first 4 time-steps 
an expansion of the time-ordered sp propagator, similarly to the method used in Ref.~\cite{Maruhn:2014}. 

\begin{table}[!htb]
\begin{tabular}{l|r|r}
\hline \hline
NEDF  &  $\lvert \delta E \rvert$ [MeV] & $\lvert \delta E / E \rvert$ (\%)\\ \hline 
SeaLL1-1  & $0.61 (0.29)$ & $0.03(0.01)$ \\
SeaLL1-2  & $0.46 (0.12)$ & $0.03(0.01)$ \\
SkM$^*$-1 & $1.03 (0.16)$ & $0.06(0.01)$ \\
SkM$^*$-2 & $0.27 (0.14)$ & $0.02(0.01)$ \\
\hline \hline
\end{tabular}
\caption{
\label{table:accuracy}
The absolute and relative error of total energy for each set of runs, and their corresponding variances in parentheses.
}
\end{table}

As mentioned above, the size of the discretized Hamiltonian in 
Eq.~\eqref{eq:tdslda} is $4 N_xN_yN_z\times 4 N_xN_yN_z$, where $N_x,\; N_y,\; N_z$ 
are the number of lattice points in the corresponding spatial directions. Each 
qpwf has 4 components and thus one has to solve $16N_xN_yN_z$ partial 
differential equations (PDEs), where each function is defined on $N_xN_yN_z$ 
lattice points. Over the years, we have developed a highly efficient code which 
takes full advantage of the graphics processing unit (GPU) accelerators 
(using compute unified device architecture (CUDA) programing) and which provides a significant 
speed-up with respect to a CPU-only code.  

The simulations were performed on Titan at OLCF, Oak Ridge, USA and Piz Daint 
in Lugano, Switzerland using a GPU code written  in CUDA. A node on Titan has 
1 GPU and 16 CPUs. A GPU code on Titan is about 9.4x faster than a CPU code 
written in C using the same number of nodes. This speed-up is practically equal 
to the theoretical limit of Titan. On Piz Daint the same GPU code is about 3x 
faster than on Titan. A fission trajectory of $\approx 3,000$ fm/c, using 512 
GPUs on Piz Daint requires less than 10 wall-time  hours, with an excellent 
strong scaling. The code was also benchmarked on Summit at OLCF, Oak Ridge, USA 
and TSUBAME, at Tokyo Institute of Technology, Japan. 

\begin{table}[!h]
\begin{tabular}{ l |  r |  r | r | r | r }
\hline \hline
Code      & CUs  & Computer     & PDEs    &  Lattice               & Cost (sec.)  \\ \hline 
Sky3D~\cite{Maruhn:2014}     &  128  & Titan             &  1,024    & $18^2\times30$  & $3.86\times10^{-6}$ \\
U\&S~\cite{Umar:2017}        &   16  & Linux cluster & 714       & $40^2\times70$  & $8.72\times10^{-5}$ \\ 
TDSLDA             &   514 & Titan             & 442,368 & $24^2\times48$  & $4.35\times10^{-8}$ \\
TDSLDA             &  240  & Piz Daint      & 442,368 &  $24^2\times48$ & $1.61\times10^{-8}$ \\
TDSLDA-opt  &  240  & Piz Daint      & 442,368 &  $24^2\times48$ & $1.23\times10^{-8}$ \\
TDSLDA             &  240  & Summit        & 442,368 &  $24^2\times48$  & $1.12\times10^{-8}$  \\
TDSLDA-opt   &  240  & Summit        & 442,368 &  $24^2\times48$  & $7.18\times10^{-9}$  \\
TDSLDA-simp    & 2     & Titan            & 684        &  $20^2\times 60$ &  $7.55\times10^{-8}$ \\
\hline \hline
\end{tabular}
\caption{\label{tab:titan}Comparison between different existing codes for 
performing TDDFT calculations on a variety of architectures. The TDSLDA code 
demonstrates an almost perfect strong scaling on Piz Daint (Lugano) and Summit 
(Oak Ridge), where further significant optimizations are likely. 
TDSLDA-opt is an optimized version of our GPU code which reduces the number of calls to
CPU-based routines.
TDSLDA-simp is a simplified and un-optimized version of our GPU code, performing the same type 
of calculations as codes~\cite{Maruhn:2014,Umar:2017} used in literature for 
TDHF+TDBCS simulations.}
\end{table}

We have compared the efficiency of our code with that of the-state-of-the-art 
codes in literature for TDHF calculations~\cite{Maruhn:2014,Umar:2017}, see 
\cref{tab:titan}. The TDHF Sky3D code~\cite{Maruhn:2014} evolves at most 
$\approx 1,000$ PDEs for the collision of two heavy-ions treating pairing 
correlations within the BCS approximation. The wall-time using a number of CPUs 
equal to the number of GPUs in our approach is almost 100x longer for similarly 
sized problems. 
It is not entirely clear how to compare codes written to solve 
somewhat different problems, TDHF and TDHFB for example. As a measure we have 
used the required computation time per lattice point of one of the components 
of a single qpwf, when performing a complete calculation of all the qpwfs 
\beq
\rm{Cost} =\frac{(\# \;\textrm{CU})\times(\textrm{wall-time})}{
(\#\; \textrm{time-steps})\times(\#\; \textrm{PDEs})\times(\#\;\textrm{lattice-points)}} ,
\eeq
where $\#$ CU stands for the number of computing units, either CPUs in case of 
the TDHF codes and GPUs in case of the TDSLDA code. We attribute the superior 
performance of the TDSLDA solver to the use of a more efficient while very 
accurate time-integration algorithm, as well as to the use of GPUs. The use of 
highly efficient and precise FFT for the computation of spatial derivatives 
could also be a factor. Since in our calculations we have to manipulate large 
amounts of data, we have taken advantage of fast I/O methods and fast 
algorithms to exchange data between computing nodes. The detailed description 
of the approach and  the code will be soon released~\cite{Jin:2019}.

\section{Saddle-to-scission times} \label{app:B}

The saddle-to-scission times 
$\tau_{s\rightarrow s}$ extracted in this round of simulations is noticeably 
shorter than those extracted in our initial study~\cite{Bulgac:2016}. 
Similar long scission-to-saddle times have been confirmed in Ref.~\cite{Tanimura:2017} 
and even longer times (up to $\approx 3\times10^4$ fm/c) were reported 
in Ref.~\cite{Simenel:2018}, in which a slight variation of the SLy4 NEDF, 
namely the SLy4d NEDF~\cite{Kim:1997}, was used. 
We  attribute these differences partially to the difference of scalar effective 
masses between the energy functionals. Indeed, for SLy4 $m*\approx 0.7m$, which 
is significantly smaller than $m*=m$ for SeaLL1 and smaller than $m*\approx 0.8m$ for SkyM*. 
One might argue that the SkM* effective mass is not much different from the 
SLy4 value. However, the Landau-Zener transition formula reads
\beq
P_{LZ}\approx
\exp \left [ - \frac{ |\Delta |^2 }{ \hbar \dot{q} \varepsilon_q(q)} \right ],
\eeq  
where the energy difference between two adjacent sp levels 
$\varepsilon_q(q) \propto 1/m*$ controls their relative slope 
with collective coordinate $q$. Since the effective mass
also determines the sp level density, even a small difference 
can lead to large changes in $P_{LZ}$. 
Another source of differences arises from the treatment of the pairing 
correlations. Relatively large variations of the scission-to-saddle times
were obtained in Ref.~\cite{Bulgac:2016} depending on whether the pairing field had 
volume, surface or mixed surface-volume character.
The magnitude of the pairing gap is typically fixed from odd-even 
nuclear mass staggering and is thus insensitive to the magnitude of the 
effective mass, which controls the sp level density at the Fermi 
level and therefore the number of sp states actively taking part 
in forming the pairing condensate. In a nucleus with a smaller effective mass, 
the size of the pairing condensate is thus diminished when compared to the 
average sp level spacing. While SLy4 describes quite reasonably gross nuclear 
properties, the sp level density at the Fermi level is drastically 
reduced when compared to observations, which are consistent with an effective 
mass $m*\approx m$. A smaller effective mass leads to a ``choppier'' potential 
energy landscape, which would inhibit the transitions at the Fermi level 
responsible for maintaining the sphericity of the local Fermi 
sphere; see Ref.~\cite{Bertsch:1980} and \cref{fig:esp_q}.

\section{Is Collective Fission Dynamics Adiabatic?} \label{app:C}

The adiabatic approximation has been the bedrock of the microscopic theory of 
fission dynamics~\cite{Schunck:2016,Pomorski:2012}
(but not of all phenomenological models) for more than half a century. 
The statement which can be found even in well established monographs~\cite{Ring:2004,Pomorski:2012} is  that the collective motion 
is so slow that one can limit the collective energy to only quadratic terms. This leads to the next conclusion, that  
the microscopic treatment of fission 
dynamics can be based on the assumption that the collective motion can be almost fully decoupled 
from the intrinsic motion. However, this assumption has never been proven to be accurate
in LACM of nuclei~\cite{Dang:2000}. 
\textcite{Ring:2004} note at the end of their Chapter 12,
where the Adiabatic Time-Dependent Hartree-Fock theory (ATDHF) is presented, that the assumption of adiabaticity 
is questionable.

Decoupling between collective and intrinsic DoF is equivalent to the assumption that 
while the collective degrees of freedom evolve in time, the intrinsics DoF remain unexcited~\cite{Ring:2004}, 
or in local thermal equilibrium in general.  
The only exception so far to this rule in literature is the 
full TDDFT~\cite{Bulgac:2013a,Bulgac:2016}, 
in which adiabaticity is not enforced.
Adiabaticity of the collective motion 
however should not be conflated with the collective dynamics having a 
quasistatic character~\cite{Schroeder:1999}. Collective motion can be very slow while 
at the same time collective and intrinsic degrees DoF are fully coupled.
Our main result is indeed that fission dynamics is most likely quasistatic in character, but not adiabatic. 
The nuclear collective 
kinetic energy at scission is an order of magnitude smaller 
that the value the nucleus would have had in the case of an adiabatic evolution.  

In a static description of the nuclear shape evolution the sp levels display an 
up and down evolution, in a manner similar to the familiar Nilsson diagrams~\cite{Ring:2004}, and they experience 
many avoided level crossings. The nucleus will retain its spherical local Fermi surface 
only if pairs of nucleons will be moved from the up-going levels to the down-going levels at these 
avoided crossings, as discussed in \cref{sec:pairing} and \cref{fig:esp_q}. 
If such transitions do not occur with unit probability the nucleus at a given nuclear shape will get excited and its 
local temperature and intrinsic entropy will increase. One might consider that a nucleus
is described by a pure wave function (in the absence of coupling to photons or 
weak interactions), since the system made of a neutron impinging 
on a heavy nucleus (subsequently fissioning) is an isolated system. 
In that case, the von Neumann entropy is constant indeed
\beq
S_\text{tot}(t)=-\text{Tr}[ \rho(t) \ln  \rho(t) ]\equiv 0. \quad \rho(t) = |\Psi (t)\rangle \langle \Psi(t) |,
\eeq
where the density matrix of the nucleus is $  \rho(t)\equiv  |\Psi(t)\rangle \langle\Psi(t)|$ 
and where $\Psi(t)$ is the exact many-body nuclear wave function 
at all times. However, what 
increases is the entanglement entropy of the intrinsic system 
\beq
S_\text{int}(t) = -\textrm{Tr}_\text{coll} [ \rho(t) \ln  \rho(t) ]\ge 0,
\eeq
where the trace is 
taken only over the collective DoF, i.e. over the ``nuclear shape'' of the fissioning nucleus. And as a result 
also the intrinsic energy and the intrinsic temperature increase as well.

\paragraph*{Adiabaticity and energy exchange.}
The intrinsic motion of the 
descending  nucleus from the outer saddle towards the scission configuration is similar to the downward motion of a 
heavy railway car on a very steep hill with its wheels blocked. The wheels dot not rotate but slip and become extremely hot due to friction, 
since almost the entire gravitational potential energy of the railway car at the top of 
the hill is converted into heat and very little of it is converted into collective kinetic energy. In this case, 
the railway car velocity is equal to the terminal velocity. An object attains a terminal velocity when the conservative 
force is balanced by the friction force, the acceleration of the system effectively vanishes and the inertia plays basically no role in its dynamics.
The motion of the railway car is strongly non-adiabatic and the velocity of the railway car is slower 
than in the case of an adiabatic evolution. However, the motion is quasistatic.

For a typical actinide the difference in the collective potential energy between the outer saddle and the scission 
configuration is $\approx 20$ MeV.  
The conserved total energy of a nucleus can always be represented in TDDFT as 
\bea
&& E_\text{tot} \!\!=
E_\text{coll}(t)+E_\text{int}(t)
\equiv \int \!\!d{\bf r}\frac{ mn({\bf r},t){\bf v}^2({\bf r},t)}{2} \nonumber \\
&& +\int \!\!d{\bf r}\, {\cal E}\left (\tau({\bf r},t)-n({\bf r},t)m^2 {\bf v}^2({\bf r},t), n({\bf r},t),...\right ), \label{eq:etota}
 \eea
where the local velocity is related to the local total momentum density ${\bf p}({\bf r},t)=m n({\bf r},t){\bf v}({\bf r},t)$.
This decomposition is possible because the TDDFT energy density satisfies local Galilean 
invariance~\cite{Engel:1975,Brink:1976,Bender:2003,Bulgac:2013a}. 
An estimate of the collective kinetic energy $E_\text{coll}(t)$ can be 
obtained also within TDGCM and ATDHF theories from the solution of the 
Schr\"odinger equation for the collective wave function. As shown by \textcite{Goeke:1980}, 
and discussed earlier by \textcite{Peierls:1962}, 
GCM with conjugate coordinates and  ATDHF 
approaches are nearly identical.

If the adiabatic approximation were valid $E_\text{coll}(t)$ would be increasing 
while the nucleus descends from saddle-to-scission  and reach a value $\approx 20$ MeV. The collective velocity 
would reach a value ${\bf v}({\bf r},t)\approx 0.01c$, which is small in comparison with the Fermi velocity $v_F\approx 0.25 c$.
In such a situation the nucleus would remain cold at all times; it would follow the lowest potential energy surface, 
the intrinsic state would be a pure state and the intrinsic entropy would be zero, since no irreversible 
energy transfer could occur between the intrinsic and collective DoF. 
The collective DoF would merely exert work on the intrinsic DoF.
At the saddle point the initial collective velocity would be zero, and the nucleus would accelerate in the descent from saddle to scission. 

\paragraph*{Overdamped evolution.}
Our results of the nuclear evolution from the saddle-to-scission however show that the nucleus experiences 
almost no acceleration: the collective inertia plays essentially no role in the dynamics. 
The collective flow energy never exceeds 1-2 MeV until the scission configuration is reached. 
This result is independent of initial conditions and NEDF used. 
The energy difference between the saddle and scission configurations is essentially entirely converted into 
intrinsic energy and never returns back to the collective degrees of freedom (DoF), which upon thermalization is converted into heat. 
Therefore, the intrinsic state of the nucleus is not a pure state anymore and the entropy of the intrinsic 
system naturally increases. The collective velocity is even smaller in magnitude now, 
$v_\text{coll}\approx 0.002\ldots 0.004 c$. In spite of the fact that the collective velocity is 
now even smaller than in an ideal adiabatic evolution, the collective motion is not more ``adiabatic.''
In thermodynamics, adiabatic processes conserve entropy and they should not be 
confused with quasistatic processes. Fission dynamics from the saddle-to-scission is a quasistatic process, 
but not an adiabatic one. As discussed in Ref.~\cite{Bulgac:2019a}, 
the inclusion of dissipation and fluctuations does not modify these conclusions.

Fission dynamics is thus similar to the overdamped motion of a Brownian particle\cite{Randrup:2011,Randrup:2011a}
\beq
m\ddot{\bf q}(t)={\bf F}({\bf q}(t))-\tensor{\gamma}\cdot \dot{{\bf q}}(t)+{\bf L}(t),
\eeq
when the acceleration $\ddot{\bf q}(t)$ vanishes this leads to 
\beq
\tensor{\gamma}\cdot \dot{{\bf q}}(t) = {\bf F}({\bf q}(t))+{\bf L}(t)
\eeq
where ${\bf F}({\bf q}(t))$ is the conservative force, $\tensor{\gamma}$ 
is the dissipation/friction tensor, and ${\bf L}(t)$ is a Langevin force.
In the absence of fluctuations a particle follows mainly the direction of the steepest descent 
\beq
\dot{\bf q}(t) = \tensor{\gamma} ^{-1}\cdot {\bf F}({\bf q}(t))
\eeq
and inertia/acceleration becomes irrelevant.
Even  in the 
case of enhanced pairing correlations, the same kind of dynamics emerges. 
In the presence of strong dissipation one cannot describe the collective dynamics 
within a Hamiltonian approach.

\paragraph*{Number of collective variables.}
The definition of collective coordinates is well known to be a problem with no satisfactory solution 
achieved yet in the microscopic theory of nuclear LACM~\cite{Dang:2000}. 
Is it possible within GCM to introduce a
parameter, which characterizes the convergence of the expansion towards a physically accurate solution, 
particularly in the case of a non-equilibrium dynamics?
Here we want to 
point to a few additional difficulties encountered while trying to introduce collective coordinates. One can ask if either TDGCM, ATDHF, or even 
the Langevin approach, are controllable approximations. In particular, can one achieve a more accurate description 
and minimize theoretical errors by increasing the number of independent collective parameters? 
Naturally, not only the needed  number of collective DoF is relevant, but also their character. In 
the case fission one typically needs at least two DoF in the initial state,  the quadrupole and octupole axial deformations
$Q_{20}$ and $Q_{30}$.
However, near the scission configuration one needs to account for the separate quadrupole and 
octupole deformations of both incipient FFs, see \cref{sec:V A,sec:V D}, and Ref.~\cite{Simenel:2018}.
 These aspects indicate that from saddle-to-scission the number of 
physically relevant DoF is likely monotonically increasing.

If the nucleus is placed on a spatial lattice with 
$N_s=N_xN_yN_z$ lattice sites, clearly the number density cannot have more than $N_s$ independent moments. Thus in 
the GCM representation where generator coordinates are restricted to moments of the density,
one cannot have more than $N_s$ independent components. In our case 
$N_s=24^2\times48=27,648$ and this would be the maximum number of independent collective coordinates possible.
 Can a fully correlated nuclear wave function be represented accurately as a sum over 
$N_s$  Slater determinants? On a spatial lattice with $N_s=27,648$ lattice sites the total number of possible 
Slater independent determinants for $^{240}$Pu (with fixed $N$ and $Z$) is
\beq
N_\text{SD}= \frac{(2N_s)!}{Z!(2N_s-Z)!}\times \frac{(2N_s)!}{N!(2N_s-N)!}\approx 10^{739}
\eeq
estimated by taking into account that at each spatial site one can place a nucleon with either spin up or down. 
While mathematically correct, this estimate is not necessarily physically correct, as most of these 
states are dynamically unreachable or physically irrelevant.  However, this estimate proves 
the point that a set of Slater determinants parameterized by all possible shapes is incomplete 
and that raises the question of whether a GCM-like parametrization is ever accurate or 
under what conditions is accurate.

There was a proposal to increase the accuracy of GCM by 
including two quasiparticle excitations~\cite{Bernard:2011}.
If a sufficient number of excitations would be introduced, one might hope 
that the total nucleus wave function is sufficiently accurate
for describing fission. If sufficiently accurate, such a description 
could account for various fluctuations not accounted for in TDDFT. 
In case of fission one has to tackle the
non-equilibrium dynamics of an open system described by the collective DoF, and the potential TDGCM representation 
of the total nucleus wave function would be
\beq
\Psi({\bf x},t) =\sum_k\int d{\bf q}  f_k({\bf q},t)\Phi_k({\bf x}|{\bf q}) ,\label{eq:TDGCM}
\eeq
where the summation over the index $k$ includes both the ground state and the 
included many-quasiparticle excited states corresponding to a fixed shape ${\bf q}$.
One typically assumes that a set of static generalized Slater determinants might be sufficient.
The question, 
which has not been addressed by \textcite{Bernard:2011} is the uniqueness of such a representation. In typical 
GCM implementations the overlap matrix
\beq
{\cal N}({\bf q},{\bf q}')  = \int d{\bf x} \Phi({\bf x}|{\bf q})\Phi({\bf x}|{\bf q}')
\eeq 
has the majority of the eigenvalues either vanishing or very small, which is a result of the strong
linear dependence of the basis set of generalized Slated determinants~\cite{Bertsch:1991}. 
In the representation \eqref{eq:TDGCM} this aspect will obviously become even worse. 
A slight change in the shape of a nucleus can always be represented as a linear superposition of 
one quasiparticle excitations. A two quasiparticle excitation can be viewed is as a part of a 
small nuclear shape change. 
Thus introducing additional quasiparticle excitations into the GCM mix clearly leads
to an additional linear dependence among basis set of states. 

One can easily convince oneself that  even with this extension of the GCM, or even of the ATDHF 
method or Langevin approach for that matter,  
the number of independent time-dependent Slater determinants would 
still be much smaller than what is needed to obtain a complete representation 
of a total nuclear wave function. According to the calculations presented in 
\cref{fig:Eq} the number of proton and neutron one quasiparticle states 
with an energy below 5 MeV is $\approx 140$ and below 
10 MeV in $^{240}$Pu is $\approx 440$, which would lead a prohibitively
large number of components in such an extension.
Moreover, it seems very unlikely that one could accurately represent 
excitation energies of the intrinsic system of $\approx 20$ MeV with only 
two quasiparticle excitations added as additional ``collective'' coordinates.''  
One might get a rough estimate by using statistical level densities~\cite{Bohr:1969, Ring:2004},
\beq
\rho(N,Z,E^*)\propto (E^*)^{-5/4}\exp\left [ 2\sqrt{aE^*}\right ], \label{eq:ldens}
\eeq
where $E^*$ is the intrinsic excitation energy and $a\approx A/10$,  or the combinatorial method~\cite{Uhrenholt:2013,Bertsch:2014}.
Already at excitation energies corresponding to an incident thermal neutron on heavy nuclei there are of the order of 
${\cal O}(10^6)$~\cite{Bohr:1969,Egidy:2005,RIPL:3} levels and at 
20 MeV excitation energy one might expect ${\cal O}(10^7)$ according to Eq.~\eqref{eq:ldens}. 
Therefore, one might rather safely  infer that a GCM or an ATDHF representation of the total wave 
function of a nucleus at excitation energies $\approx 20$ MeV would be numerically intractable, unless restricted to  
a few collective coordinates. Naturally, one may ask how many independent Slater determinants are needed to reach a given accuracy.  
This still remains an open question. 

Some of the present authors believe that an
improvement over current implementations of the TDGCM, ATDHF, and Langevin approaches 
is the approach outlined in~\cite{Bulgac:2019a}, in which all collective 
DoF are taken into account, plus dissipation and fluctuations, in a fully quantum extension of TDDFT.  
Would such an 
approach be theoretically sufficiently accurate? So far the description of mass yields using various implementations
of the Langevin approach~\cite{Frobrich:1998,Randrup:2011,Randrup:2011a,
Sierk:2017,Ishizuka:2017,Sadhukhan:2016,Sadhukhan:2017} leads apparently in all cases to qualitatively similar results, 
in satisfactory agreement with experiment. These results might simply point 
to the fact that these observables are relatively weakly sensitive 
to details of the Langevin approach implementation. 



\section{Critique of the stochastic mean field prescription} \label{app:E}

One can attempt to simulate the effect of a superposition of 
(generalized) Slater determinants suggested by the path-integral approach 
by following the stochastic mean field model 
introduced by Ayik~\cite{Ayik:2008}.  In the stochastic mean field model 
\emph{fluctuations only stem from the fluctuations in the initial density}~\cite{Tanimura:2017}
and the time evolution is exactly the usual time-dependent mean field. 
This ad hoc assumption 
is at odds with the Langevin approach and also with the path-integral 
approach, in which fluctuations along the entire path are relevant. 

One can easily check that 
in any classical Langevin description, if fluctuations vanish after a certain finite time, friction
erases their memory in the long time limit. For example, if the force is constant, the 
solution of the classical Langevin equation  is
\bea  
&& \!\!\! \!\!\! m\dot{v}(t)= F-\gamma m v(t) +m\xi(t), \label{eq:l1}\\
&& \!\!\! \!\!\! v(t) =v(0)e^{-\gamma t} + \frac{F}{m\gamma}(1-e^{-\gamma t}) +e^{-\gamma {t}}\int_0^{t}\!\!\! \!\!\! dt' \xi(t')e^{-\gamma t'}, \label{eq:l3}
\eea
and the integral term becomes a constant and the role of fluctuations soon becomes exponentially small  
and the particle continues moving with a constant terminal velocity.

The conclusion reached in Ref.~\cite{Tanimura:2017}  is at odds also 
with our quite firm conclusion that in nuclear LACM
the memory of the initial conditions is washed out 
during the evolution rather quickly, see \cref{sec:V A}, \cref{sec:V C}, and \cref{sec:V D}.
In our approach we follow 
basically a similar strategy, the trajectories are obtained from a time-dependent mean field dynamics.
We chose  our initial conditions from an ensemble of
initial energies and initial collective variables with similar spreads.
Because the one-body dissipation is so effective in bringing the 
collective flow almost to a stop, at any point on the potential energy surface 
the system will most likely follow the direction of the steepest descent, see \cref{app:B},  and 
the collective inertia will have a marginal effect on the collective dynamics. 
Therefore, in its evolution from saddle-to-scission, the nucleus will largely 
follow the bottom of the fission valley. The collective nuclear motion becomes 
very similar to the motion of a viscous fluid. 

Ayik's model is phenomenological in nature, like random matrix 
theory, since it involves simulating quantal fluctuations of observables with a 
random ensemble. This makes the statement of \textcite{Tanimura:2017} that they 
obtained for the ``first time a fully microscopic description of the fragment 
TKE distribution after fission'' questionable.

We suggest that the discrepancy between the results of Ref.~\cite{Tanimura:2017} 
and the path-integral approach, the Langevin approach, and our results too 
arises from the large unphysical
fluctuations of all physical observables inherent to the stochastic mean field 
approach, the nature of which we describe below.
Since in the stochastic mean field method \emph{fluctuations only stem from the
fluctuations in the initial density}~\cite{Tanimura:2017} one would expect that their conclusions should parallel ours, 
as we have considered  a relatively large set of initial conditions with a similar spread in initial energies 
and deformations.  

\begin{figure}
\includegraphics[width=\columnwidth]{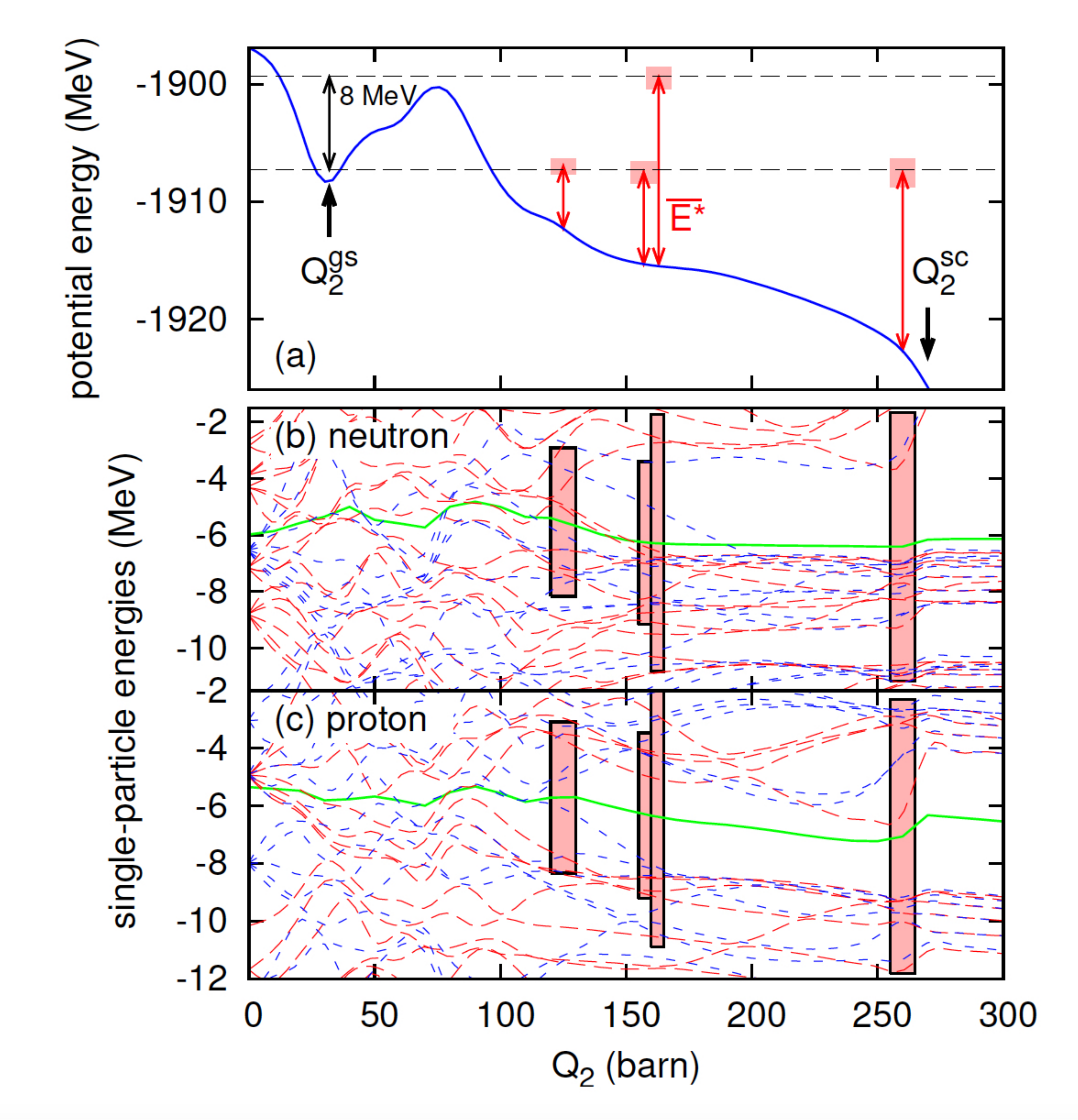}
\caption{\label{fig:tanimura} (Color online)
Occupation probabilities of the nucleon levels within a window of approximately 
5 MeV were prescribed random values as in Eqs. \eqref{eq:xi}, \eqref{eq:tdhf}, 
and \eqref{eq:sigma}. The red bars show the size of the sp energy 
window, in case of each deformation, where stochastic fluctuations are allowed. 
This is a copy of the Fig. 1 from the Supplemental Material of 
Ref.~\cite{Tanimura:2017}.}
\end{figure}

In the stochastic mean field prescription one uses an ensemble 
$\{ \lambda\}_{\lambda\in\Lambda}$ of one-body density matrices 
$\rho^{\lambda}_{kl}$ such that 
\bea 
\rho^{\lambda}({\bm r},{\bm r'},t) & = &\sum_{k,l} \phi_k({\bm r},t)\rho^{\lambda}_{kl}\phi^*_l({\bm r'},t), 
\label{eq:xi}\\
\rho^{\lambda}_{kl} & = & \rho^{\lambda *}_{lk} = n_k\delta_{kl} + \xi^{\lambda}_{kl}, 
\label{eq:tdhf} \\
i\hbar\dot{\phi}_k({\bm r},t) & = & h[\rho^{\lambda}]\phi_k({\bm r},t)
\eea
where $n_k$ are initial time-independent, zero-temperature, sp 
occupation probabilities obtained by considering pairing interactions, 
$\xi^{\lambda}_{kl}= \xi^{\lambda *}_{lk}$ are time-independent, independent 
Gaussian complex random numbers with zero mean and variance
\bea
\sigma^2_{kl} 
= \overline{\xi^{\lambda}_{kl}\xi^{\lambda *}_{kl}} 
= \frac{1}{2}\left [ n_k(1-n_l) + n_l(1-n_k) \right ], 
\label{eq:sigma} 
\eea
where overline refers to statistical averaging over the 
events $\lambda$ of the ensemble $\Lambda$. All other second moments of the 
distributions vanish. These Gaussian random numbers are chosen non-vanishing in 
a limited energy window around the Fermi level, see \cref{fig:tanimura}.
By allowing these random fluctuations in the one-body density matrix the total 
energy of the system also fluctuates. Both the intrinsic excitation energy of 
the nucleus and the size of the fluctuations of the total energy are controlled 
by the size of the sp energy window where fluctuations are 
non-vanishing.

A particularly illuminating example, which illustrates the difficulties of 
the stochastic mean field model is the case of a single particle, when $(n_0=1, \; n_{k>0}=0$)
\bea
&\rho^{\lambda}({\bm r},{\bm r'},t) = \phi_0({\bm r},t)\phi^*_0({\bm r'},t) +\\
&\sum_{k=1}^M\left [ \xi_{0k}^{\lambda}  \phi_0({\bm r},t)\phi^*_k({\bm r'},t)+
\xi_{0k}^{\lambda *}  \phi_k({\bm r},t)\phi^*_0({\bm r'},t)\right ],\nonumber \\
&\overline{\xi^{\lambda}_{0k}\xi_{0l}^{\lambda*}}=\frac{1}{2}\delta_{kl}, \quad \overline{\xi^{\lambda}_{0k}}=0.
\eea
Simple algebraic manipulations show that the eigenvalue equation
\beq
\int d{\bf r}' \rho^{\lambda}({\bm r},{\bm r'},t)\psi({\bf r}',t)  = \nu \psi({\bf r},t)
\eeq
has two non-vanishing eigenvalues
\bea
\nu_{1,2} &= \frac{1}{2}\pm \sqrt{ \frac{1}{4}+ \sum_{k=1}^M |\xi_k|^2}\approx \frac{1}{2}\pm\sqrt{\frac{1}{4}+\frac{M}{2}},\\
& \lim_{M\rightarrow\infty}\nu_{1,2} = \pm\infty,
\eea
even though $\nu_1+\nu_2\equiv 1$. Such a stochastic fermionic density matrix
violates blatantly the Pauli principle. One can also show that
\bea
&&\int d{\bf r}'' \rho^{\lambda}({\bm r},{\bm r''},t)\rho^{\lambda}({\bm r''},{\bm r'},t)=
\rho^{\lambda}({\bm r},{\bm r'},t)  \\
&&+ \sum_{k=1}^M \phi_0({\bf r},t)\xi_k^*\phi_k^*({\bf r'},t) 
  + \sum_{k=1}^M \xi_k\phi_k({\bf r'},t) \phi_0^* ({\bf r'},t) \nonumber \\
&&+ \phi_0({\bf r},t)\phi_0^*({\bf r'},t) \sum_{k=1}^M|\xi_k|^2 + 
\sum_{k=1}^M\xi_k\phi_k({\bf r},t) \sum_{l=1}^M\xi_l^*\phi_l^*({\bf r'},t), \nonumber \\ 
&& \overline{\Tr \rho^\lambda } =0,\quad \overline{\Tr [\rho^\lambda ]^2} = 1 + M,\quad \lim _{M\rightarrow \infty} \Tr [\rho^\lambda ]^2= \infty .
\eea
Even though the particle number is conserved on average in the stochastic mean field approach,
the variance of the particle number is extremely large, or even infinite if the entire sp
spectrum is included. These higher moments enter into the expressions of the total energy and the
sp Hamiltonian.

\begin{figure}[!htb]
\includegraphics[width=\columnwidth]{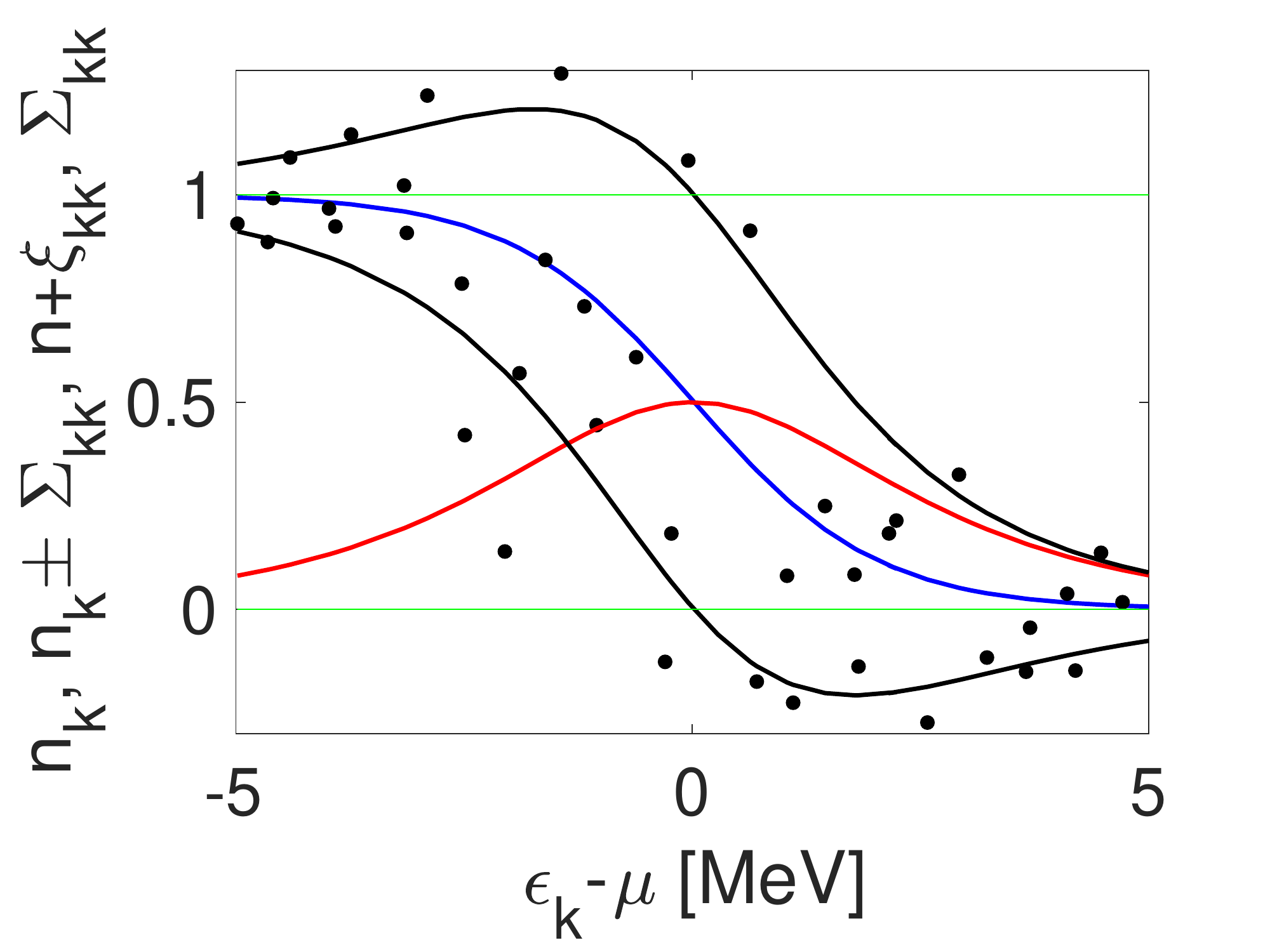}
\caption{\label{fig:occup} (Color online)
Average neutron occupation probabilities $n_k$ (blue) for a system with 
$N=150$, a Fermi energy $\varepsilon_F$ = 35 MeV, and an almost constant 
average sp level density at a temperature 1 MeV, the 
$\rho_{kk}=n_k \pm  \sigma_{kk}$ (black), $\sigma_{kk}$ (red), and a typical 
random realization of the stochastic occupation probabilities $n_k+\xi_{kk}$ 
(black dots) chosen in an energy window (-5, 5) MeV around the Fermi level.}
\end{figure}

For an arbitrary number of particles the Hermitian matrix $n_k\delta_{kl}+\xi^{\lambda}_{kl} = \sum_m 
S^{\lambda}_{km} \nu^{\lambda}_{m} S^{\lambda *}_{lm}$ can be diagonalized and 
the density matrix can be re-written as
\beq 
\label{eq:nu}
\rho^{\lambda}({\bm r},{\bm r'},t) 
= \sum_m \psi^{\lambda}_m({\bm r},t) \nu^{\lambda}_{m}\psi^{\lambda *}_m({\bm r'},t) 
\eeq
where $\psi^{\lambda}_k({\bm r},t) = \sum_l \phi_l({\bm r},t)S^{\lambda}_{lk}$ 
and $\sum_m S^{\lambda}_{km}S^{\lambda *}_{lm}=\delta_{kl}$. Both sets of
sp wave functions in Eq. \eqref{eq:xi} or in the equivalent Eq. 
\eqref{eq:nu} satisfy the same TDHF equations \eqref{eq:tdhf} with the 
sp Hamiltonian $h[\rho^{\lambda}]$ defined through the random 
density matrix $\rho^{\lambda}({\bm r},{\bm r'},t)$ Eq. \eqref{eq:xi}.

\begin{figure}[!htb]
\includegraphics[width=\columnwidth]{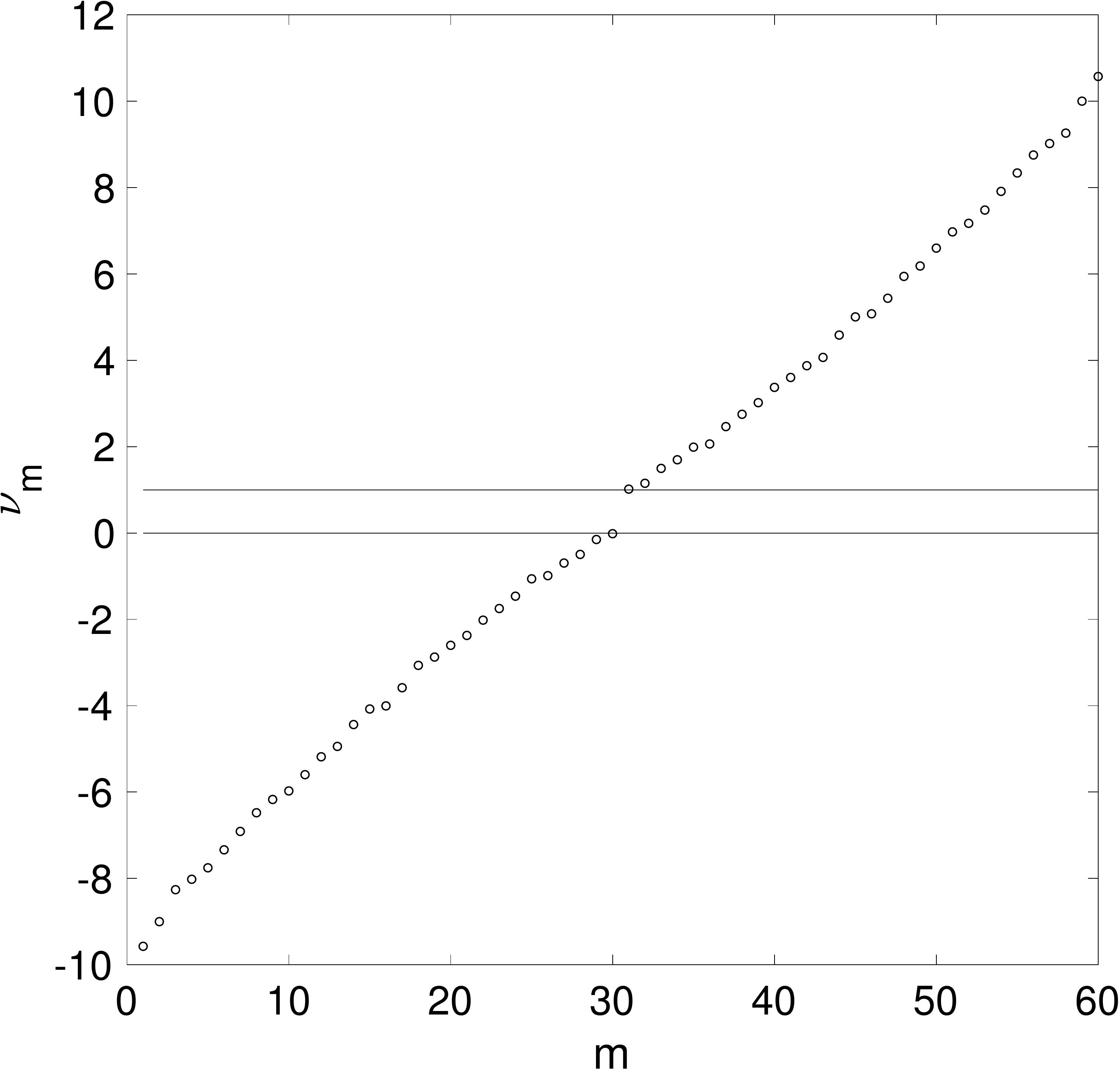}
\caption{\label{fig:occup1} the ``occupation probabilities'' $\nu_m$ \eqref{eq:nu} after the diagonalization of 
one stochastic realization of Eq. \eqref{eq:tdhf} for $n_{k=1\ldots 30} =1$ and $n_{l=31\ldots 60}=0$.
Two horizontal lines are at 
 0 and 1 levels.  The shape of this spectrum of the ``occupation probabilities'' changes 
little from one stochastic realization to another. The range of $\nu_m$ values 
increases with the size of the energy interval over which fluctuations are 
allowed, covering the entire real axis, if the fluctuations are 
allowed over the entire sp spectrum.   }
\end{figure}

In \cref{fig:occup} and \cref{fig:occup1} we illustrate the 
problems with the stochastic mean field prescription, which leads to effective 
occupation probabilities $\nu^{\lambda}_m$ outside the physical interval [0, 1], 
which is a flagrant violation of the Pauli principle.
The statistical average value of the particle number is correct
\bea
\overline{\langle \hat{N}\rangle}
= \overline{\text{tr}\rho^{\lambda}} 
= \sum_k \overline{n_k+\xi^{\lambda}_{kk}}
= \sum_k n_k.
\eea

However, since the fluctuations are very large (even infinite if the entire spectrum is included) 
the total energy of the nucleus has also very large unphysical fluctuations.
Let us estimate now the statistical average 
of a typical interaction term 
\bea
&& \int d{\bm r} \overline{\rho^2({\bm r},{\bm r},t)}
= \int d{\bm r} \left[ \sum_k n_k|\phi_k({\bm r},t)|^2\right ] ^2 \\
&& +\frac{1}{2} \int d{\bm r} \sum_{kl}  |\phi_k({\bm r},t)|^2|\phi_l({\bm r},t)|^2[n_k+n_l-2n_kn_l] \nonumber \\
&& =\int d{\bm r} \sum_{kl}n_k   |\phi_k({\bm r},t)|^2 |\phi_l({\bm r},t)|^2\approx \frac{N N_{\rm lev}}{V}, \nonumber 
\eea
where we have considered for the sake of simplicity of the argument only the sp states 
in the window where fluctuations are allowed.
$V$ is the volume of the system.
This can be very different from the real  value
\beq
\int d{\bm r} \rho^2({\bm r},{\bm r},t) 
 \approx N \frac{N}{V}\ll  N_{\rm lev} \frac{N }{V}.
\eeq
In Ref. \cite{Tanimura:2017} the authors use the parameter $N_{\rm lev}> N$ to 
control by how much they can affect the internal energy of the nucleus. A 
simpler and equally arbitrary approach would be to multiply the density 
$\rho({\bm r},{\bm r},t)$ by $1+\xi$, where $\xi$ is a random number with zero 
mean. The size of the fluctuations of the energy of the nucleus is 
controlled in Ref.~\cite{Tanimura:2017} by the arbitrary number of 
the sp levels $N_{\rm lev}$, or by the arbitrary size of the 
sp energy window where such fluctuations are allowed. As a 
result the unphysical particle fluctuations lead also to unphysical energy fluctuations
in the stochastic mean field approach.

There is no theoretical argument presented in Ref.~\cite{Tanimura:2017} on how to 
choose the starting point of the dynamical simulations.  For $Q_{20}> 125$ barn, as is clearly 
seen from \cref{fig:tanimura}, the FFs have been already well 
individualized and no redistribution of the sp occupation numbers 
occurs anymore. At these deformations the size of this sp energy 
window is chosen so as to reproduce on average the increase in the total energy 
of the nucleus to match the ground state energy. It would seem more natural to 
start the simulation at the exact configuration where the nucleus emerged from 
under the barrier at $Q_{20}\approx$ 100 barn. In that case the size of the 
``fluctuations'' would be zero, as the nucleus emerges from under the barrier 
in its intrinsic ground state. Starting at the deformation $Q_{20}\approx$ 100 
barn, however, would deprive the authors from the ability to generate a desired 
FFs distribution, induced by the presence of fluctuations. The authors even 
establish that if they were to start their simulations closer to the scission 
configurations their results would be quite different, thus precluding this 
approach of its predictive power. One can thus safely conclude that the FFs 
distributions definitely depend strongly on the choice of the initial 
conditions within such a stochastic mean field approach. 


\providecommand{\selectlanguage}[1]{}
\renewcommand{\selectlanguage}[1]{}

\bibliography{local_fission}

\end{document}